\documentclass[twocolumn,amsmath,aps,fleqn]{revtex4}

\usepackage{amssymb}
\usepackage{amsmath} 
\usepackage{epsfig}
\usepackage{subfigure}
\usepackage{mathrsfs}
\usepackage{longtable}

\newcommand{\be}{\begin{eqnarray}}
\newcommand{\ee}{\end{eqnarray}}
\newcommand{\bea}{\begin{eqnarray}}
\newcommand{\eea}{\end{eqnarray}}

\begin{document}

\title{Towards a fully consistent parameterization of modified gravity}
\author{Tessa Baker}
\email{tessa.baker@astro.ox.ac.uk}
\affiliation{Astrophysics, University of Oxford, DWB, Keble Road, Oxford, OX1 3RH, UK}
\author{Pedro G. Ferreira}
\email{p.ferreira1@physics.ox.ac.uk}
\affiliation{ Astrophysics and Oxford Martin  School, University of Oxford, DWB, Keble Road, Oxford, OX1 3RH, UK}
\author{Constantinos Skordis}
\email{skordis@nottingham.ac.uk}
\affiliation{School of Physics and Astronomy, University of Nottingham, University Park, Nottingham,  NG7  2RD,UK}
\author{Joe Zuntz}
\email{jaz@astro.ox.ac.uk}
\affiliation{Astrophysics and Oxford Martin  School, University of Oxford, DWB, Keble Road, Oxford, OX1 3RH, UK\\
Dept. of Physics \& Astronomy, University College London, WC1E 6BT, UK}

\begin{abstract}
There is a distinct possibility that current and future cosmological data can be used to constrain Einstein's theory of gravity on the very largest scales. To be able to do this in a  model-independent way, it
makes sense to work with a general parameterization of modified gravity. Such an approach would
be analogous to the Parameterized Post-Newtonian (PPN) approach which is used on the scale of the Solar System. A few such parameterizations have been proposed and preliminary constraints have been
obtained. We show that the majority of such parameterizations are only exactly applicable in the quasistatic regime. On larger scales they fail to encapsulate the full behaviour of typical models currently under consideration.
We suggest that it may be possible to capture the additions to the `Parameterized Post-Friedmann' (PPF) formalism by treating them akin to fluid perturbations.

\end{abstract}

\maketitle
\section{Introduction}
\label{Intro}

It is possible that we live in a Universe in which more than 96$\%$ of the energy
and matter density is in the form of an exotic dark substance. The conventional
view is that roughly a quarter of this obscure substance is in the form of dark matter
and the remainder is in the form of dark energy. Theories abound that propose explanations
for dark matter and dark energy and there is an active programme of research
attempting to understand and measure them.

It may also be possible that our understanding of gravity is lacking, and that Einstein's
theory of General Relativity (and more specifically, the Einstein field equations) are
not entirely applicable on cosmological scales. The past decade has seen unprecedented growth,
from a handful to a veritable menagerie of possible modifications to  gravity that
may be perceived as a fictitious dark sector \cite{MG_report}.

The proliferation of theories of modified gravity couldn't have come at a better time. Observational
cosmology has entered what some have called an era of `precision cosmology'. Hubristic
as such a point of view might be, it is certainly true that cosmology is being inundated by
data, from measurements of the Cosmic Microwave Background (CMB) \cite{WMAP, Planck}, galaxy surveys \cite{SDSS}, weak lensing surveys \cite{DES} and measurements of distance and luminosity at high redshift with supernovae Ia \cite{Riess2011}. With such
data in hand it is possible to test cosmological models and constrain their parameters with some
precision. With the forthcoming experiments currently on the drawing board \cite{SKA, EIC, LSST}, great things are expected. In particular, there is a hope that it may be possible to distinguish between the two paradigms: the dark sector versus modified gravity.

The situation in cosmology is reminiscent of that in General Relativity in the late 1960's
and early 1970's. Then, Einstein's theory was undergoing a golden age with discoveries
in radio and X-ray astronomy, as well as precision measurements in the Solar System and beyond, making it increasingly relevant. As a result, a plethora of
alternative theories of gravity were proposed which could all in principle be tested (and ruled out) by
observations \cite{Will_linear_disproof, Will_Nordvedt_preferred_frame,Will_living_review_2006}. Out of this situation a phenomenological model of modified gravity emerged, the Parameterized Post-Newtonian (PPN) approximation \cite{Will1971, Thorne_Will, Will_Nordvedt_1972, Will2011}, which could be used as a bridge between theory
and observations. In other words, from observations it is possible to find the constraints on the parameters
in the PPN approximation. The constraints are model-independent. From any given theory it is
then possible to calculate the corresponding PPN parameters and find if they conform to observations.
The PPN approximation is sufficiently general that it can encompass almost all modified theories
of gravity that were then proposed.

Clearly something like the PPN approach is desirable in cosmology. Given the rapid increase
in the number of modified theories of gravity, it would make sense to construct a parameterization that
could serve as a bridge between theory and observation. Observers could express their constraints
in terms of a set of convenient parameters; theorists could then make predictions for these parameters and check if their theories are observationally viable. 
Instead of performing many constraint analyses on individual theories, one could run just a single constraint analysis on the parameterized framework. With a dictionary of translations between theories and the parameterization in hand, these general constraints could be immediately applied to any particular theory. 

Another key advantage of a parameterized approach is that it allows one to explore regions of theory space for which the underlying action is not known. For example, in \textsection\ref{subsection:f_r} we will see how the Lagrangian $f(R)$ is related to our framework  `parameters'  (which are really functions, not single numbers -- see below). Cosmological data will exclude certain regions of parameter space. If a new form of $f(R)$ is proposed in the future, it should be a quick operation to see whether it falls into the excluded region - even though that particular Lagrangian was not known at the time that the constraint analysis was performed.

In this paper we will discuss the requirements of how to parametrize modifications to gravity on cosmological scales. It develops the principles first put forward in \cite{Skordis} and explores
how they may be applied more generally. The layout of this paper is as follows. In \textsection\ref{section:conventional} we discuss the idea behind the PPN approach and show that it can't
be imported wholesale into cosmology. We then briefly look at attempts at parametrizing gravity and point out their limitations. In \textsection\ref{section:formalism} we discuss a possible formalism in detail -- we co-opt the name Parameterized Post-Friedmann approach from \cite{Hu_Sawicki} -- and argue that it may be sufficiently general to encompass a broad class of theories. In \textsection\ref{section:no_extra_fields} we construct the hierarchy of equations that should be satisfied in the case in which there
are no extra fields contributing to the modifications to gravity. In \textsection\ref{section:extra_fields} we discuss the more general case {\it with} extra fields and how this affects the relations between the different coefficients. In \textsection\ref{section:examples} we focus on four modified theories of gravity and
analyse how they fit into the formalism that we are proposing. Finally, in \textsection\ref{section:discussion} we summarize the state of play of the parameterization we are proposing.
\section{The conventional approach}
\label{section:conventional}
The plan is to construct a parameterization that might mimic the PPN approach on cosmological
scales. It is therefore useful to look very briefly at the PPN approach, which proceeds as a perturbative expansion in $v/c$ (though we set c=1 in what follows). Consider the modified, linearized Schwarzschild solution: 
\begin{eqnarray}
ds^2&=&-\left(1-2\Psi\right)dt^2+\left(1-2\Phi\right)dr^2+r^2d\Omega^2 \nonumber \\
\Psi&=&\frac{G_0M}{r} \nonumber \\
\Phi&=&\gamma_{\tt PPN} \Psi
\end{eqnarray}
where $G_0$ is Newton's constant, $M$ is the central mass, $\Omega$ is the two-dimensional volume element of a sphere, ($\Psi$, $\Phi$) are gravitational potentials and $\gamma_{\tt PPN}$ is a PPN parameter, equivalent to one of the older Eddington-Robertson-Schiff parameters.

There are a few properties which are of note in this expression. First of all, the parameterization is
constructed around a {\it solution} of the Einstein field equations, the Schwarzschild solution
(with \mbox{$\gamma_{\tt PPN}=1$}) -- the field equations do not come into play. Second (and this isn't obvious from the expressions above), the parameter $\gamma_{\tt PPN}$ only depends on parameters in the theory and not on integration
constants or `environmental' parameters such as the central mass. This means that, given a theory, it is possible to predict $\gamma_{\tt PPN}$ solely in terms of fundamental parameters of the theory (i.e. the parameters in the action). Finally, we see that the mismatch between the gravitational potentials can be expressed as \cite{Thorne_Will}
\begin{eqnarray}
\Phi-\Psi=\zeta_{\tt PPN}\Phi \label{PPNslip}  
\end{eqnarray}
with $\zeta_{\tt PPN}=(\gamma_{\tt PPN}-1)/\gamma_{\tt PPN}$ often called the {\it gravitational slip}. General Relativity is recovered when $\zeta_{\tt PPN}=0$.

The idea of applying an equation of the form of eqn.(\ref{PPNslip}) to cosmology emerged from the work
of Bertschinger in \cite{Bertschinger2006}. Bertschinger showed that on large scales it was possible to calculate the evolution of $\Psi$ and $\Phi$ using only information about the background
evolution and assuming a closure relationship between the two potentials. The simplest assumption
is a closure relation of the form of eqn.(\ref{PPNslip}), but in no way was it implied that this would be
a realistic relationship that would be valid in the general space of theories of modified gravity. 

Nevertheless, over the last few years the simplified equation for gravitational slip has been adopted as a general parameterization which should be valid in cosmology \cite{Pogosian_parameterization, Bean, Daniel2010}. It has been shown to be valid in a few cases, in the quasistatic regime (i.e. on small scales), and explicit expressions have been found for
$\zeta$ in terms of fundamental parameters of those theories (some examples are collected in \cite{MG_report}). Such a parameterization has been extended to include another parameter, a modified Newton's constant $G_{\rm eff}$, which may differ from $G_0$. The method is then to use eqn.(\ref{PPNslip}) and a modified Poisson equation,
\begin{equation}
-k^2\Phi = 4\pi G_{\rm eff} a^2 \sum_i \rho_i \Delta_i 
\label{convpoisson} 
\end{equation}
(where $\rho_i$ is the energy density of fluid $i$ and $\Delta_i$ is the comoving energy density) to modify the evolution equations for cosmological perturbations. A modified Einstein-Boltzman solver is then used to calculate cosmological observables.  
The two parameters ($G_{\rm eff}$, $\zeta$) have been adopted more generally and have been used to find preliminary constraints on modified gravity theories by a number of groups \cite{Bean, Lombriser2010,Daniel2010, Daniel_Linder}.

Clearly such an approach to parametrizing modified gravity has some significant differences with the
PPN approach. For a start, modifications are applied to the field equations and {\it not} to specific solutions of Einstein's field equations. This is understandable -- the solutions of interest in cosmology are not only inhomogeneous but time-varying, unlike the incredible simplicity of the Schwarzschild solution that arises due to Birkhoff's theorem. Also, unlike in PPN, the parameters at play -- $\zeta$ and $G_{\rm eff}$ -- will not only depend on fundamental parameters of the theory but also on the time evolution of the cosmological background.

Ideally, any time-dependence in the parameterization will be simply related to background cosmological
quantities (like the scale factor, energy densities or any auxiliary fields that are part of the modifications)
and {\it not} dependent upon the time evolution of $\Phi$, $\Psi$ or any other perturbation variables. For such a requirement to be possible it is essential that any parameterization is sufficiently general to encompass
a broad range of theories. As we will show in this paper, parameterizations using eqns.(\ref{PPNslip}) and (\ref{convpoisson}) are simply not general enough to capture the full range of behaviour of modified theories of gravity. 
 It has been argued that such a parameterization can be used as a diagnostic; that is, for example, a non-zero measurement of $\zeta$ might indicate modifications of gravity \cite{Hojjati_Pogosian}. This may be true, but such a measurement cannot then be used to go further and constrain
specific theories. It would be more useful to build a fully consistent parameterization which can be used as a diagnostic {\it and} can be linked to theoretical proposals. The purpose of this paper is to take the first steps towards such a parameterization.
\section{The Formalism}
\label{section:formalism}
When considering modified gravity theories it can be helpful to cleanly separate the non-standard parts from the familiar terms that arise in General Relativity (henceforth GR). We can always write the modifications as a an additional tensor appearing in the Einstein field equations, i.e.
\begin{equation}
\label{einstein}
G_{\mu\nu} \;=\; 8\pi G_0a^2 T_{\mu\nu}+a^2U_{\mu\nu}
\end{equation}
The diagonal components of the tensor $U_{\mu\nu}$ are equivalent to an effective dark fluid with energy density $X$ and isotropic pressure $Y$ (where the constants have been absorbed). The zeroth-order Einstein equations are then:
\begin{eqnarray} 
E_F &\equiv&3{\cal H}^2+3 K=8\pi G_0 a^2 \sum_i \rho_i +a^2 X \label{E_F} \\ 
E_R&\equiv&-(2\dot{\cal H}+{\cal H}^2+K)=8\pi G_0 a^2 \sum_i P_i +a^2 Y \label{E_R} 
\end{eqnarray}
where ${\cal H}=H/a$ is the conformal Hubble parameter and $K$ is the curvature. We will use $E_F$ and $E_R$ as defined above throughout this paper.  For future use we define $E=E_F+E_R$. The summations in the above expressions are taken over all conventional fluids and dark matter, and dots denote derivatives with respect to conformal time. In this paper we will largely adhere to the definitions and conventions used in \cite{Skordis}. 
We also note that the Bianchi identity $\nabla_{\mu}G^{\mu}_{\nu}=0$ implies the relation:
\begin{equation}
\label{E_Bianchi}
\dot E_F+{\cal H}(E_F+3 E_R)=0
\end{equation}
Assuming that the conservation law $\nabla_{\mu}T^{\mu}_{\nu}=0$ holds separately for ordinary matter and the effective dark fluid,  $X$ and $Y$ must be related by the equation $\dot X+3 {\cal H}(X+Y)=0$. 

Continuing in this vein, our goal is to write the linearly perturbed Einstein equations as:
\begin{equation}
\label{einstein_perturbed}
\delta G_{\mu\nu} \;=\; 8\pi G_0a^2\delta T_{\mu\nu}+a^2\delta U_{\mu\nu}
\end{equation}
In general the tensor $\delta U_{\mu\nu}$ will contain both metric perturbations and extra degrees of freedom (hereafter d.o.f) introduced by a theory of  modified gravity. We can separate $\delta U_{\mu\nu}$ into three parts: i) a part containing only metric perturbations, ii) a part containing perturbations to the extra d.o.f., iii) a part mixing the extra d.o.f. and perturbations to the ordinary matter components: 
\begin{equation}
\label{U_decomposition}
\delta U_{\mu\nu} =\delta U_{\mu\nu}^{\textrm{metric}}(\hat\Phi, \hat\Gamma...)+\delta U_{\mu\nu}^{\mathrm{dof}}(\chi, \dot\chi, \ddot\chi...)+\delta U_{\mu\nu}^{\textrm{mix}}(\delta\rho...) 
\end{equation}
The argument variables in this expression will be introduced shortly. 
We have written the Einstein field equations such that $T^{\mu}_{\nu}$ contains only standard, uncoupled matter terms and hence obeys the usual (perturbed) conservation equations, \mbox{$\delta (\nabla_{\mu}T_{\nu}^{\mu})=0$}. As a result $U^{\mu}_{\nu}$ must obey its own independent conservation equations, so that at linear order we have \mbox{$\delta (\nabla_{\mu} U^{\mu}_{\nu})=0$}. We will use the following notation to denote components of $\delta U^{\mu}_{\nu}$ from here onwards: 
\begin{eqnarray}
\label{U_components_def}
U_{\Delta}&=&-a^2\delta U^0_0, \qquad  \vec{\nabla}_i U_{\Theta}=-a^2 \delta U^0_i\\
 U_P&=&a^2\delta U^i_i, \qquad\,\, D_{ij}U_{\Sigma}=a^2(U^i_j-\frac{1}{3}\delta U^k_k\delta^i_j)\nonumber
\end{eqnarray}
where $D_{ij}=\vec{\nabla}_i\vec{\nabla}_j-1/3 q_{ij}\vec{\nabla}^2$  projects out the longitudinal, traceless part of $\delta U^{\mu}_{\nu}$ 
and $q_{ij}$ is a maximally symmetric metric of constant curvature $K$. The definition of $U_{\Sigma}$. In the case of unmodified background equations perturbed conservation equation for $U_{\nu}^{\mu}$ gives us the following two constraint equations at the linearized level \cite{Skordis}:
\begin{eqnarray}
\label{Bianchi1}
\dot U_{\Delta}+{\cal H} U_{\Delta}-\vec{\nabla}^2 U_{\Theta}+\frac{1}{2}a^2 (X+Y)(\dot\beta+2\vec{\nabla}^2\epsilon) \nonumber \\+{\cal H}U_P &=&0 \\
\dot U_{\Theta}+2{\cal H} U_{\Theta}-\frac{1}{3}U_P-\frac{2}{3}\vec{\nabla}^2 U_{\Sigma}+a^2(X+Y)\Xi &=&0 
\label{Bianchi2}
\end{eqnarray}
where the metric fluctuations $\beta,\,\epsilon$ and $\Xi$ are defined in eqn.(\ref{line_element}).

In this paper we will initially present general forms for the construction of metric-only $\delta U_{\mu\nu}$ that satisfy equations (\ref{Bianchi1}) and (\ref{Bianchi2}), then impose the restriction that the field equations can contain at most second-order derivatives. Gravitational theories containing derivatives greater than second-order are generally disfavoured as they typically result in instabilities or the presence of ghost solutions \cite{MG_report, Chiba_ghost, Ostrogradsky}. However, we note that some special cases of higher-order theories are acceptable e.g. $f(R)$ theories \cite{Woodard} (see \textsection\ref{subsection:f_r}). Hence we start with the general case in order to indicate how our results may be extended to higher-derivative theories \cite{Stelle}.

The requirement of second-order field equations means that $U_{\Delta}$ and $U_{\Theta}$ can only contain first-order derivatives with respect to conformal time, as can be seen from eqns.(\ref{Bianchi1}) and (\ref{Bianchi2}). The specific implications this has depends on which of the tensors in eqn.(\ref{U_decomposition}) are present. In \textsection\ref{section:no_extra_fields} we will explore the structure of theories with only metric perturbations, whilst theories with extra degrees of freedom will be presented in \textsection\ref{section:extra_fields}. In Appendix \ref{app:constraints} we display formulae for generating constraint equations in an arbitrary-order theory of gravity with no additional degrees of freedom.

We write the perturbed line element (for scalar perturbations only) as: 
\begin{eqnarray}
\label{line_element}
ds^2 = &-&a^2 (1-2 \Xi)dt^2-2a^2\,(\vec{\nabla}_i\epsilon)dt\,dx\nonumber\\&+&a^2\left[\left(1+\frac{1}{3}\beta\right)q_{ij}+D_{ij}\nu\right] dx^i\,dx^j 
\end{eqnarray} 
It will prove useful to define the gauge-variant combination:
\begin{equation}
 V=\dot\nu+2\epsilon 
\end{equation}
We also define the gauge-invariant potentials:
\begin{eqnarray}
\hat\Phi &=& -\frac{1}{6}(\beta-\vec{\nabla}^2\nu)+\frac{1}{2}{\cal H}V \label{phi_def}\\
\hat\Psi &=& -\Xi-\frac{1}{2}\dot V-\frac{1}{2}{\cal H}V \label{psi_def}
\end{eqnarray}
$\hat\Phi$ and $\hat\Psi $ are equivalent to the Bardeen potentials $-\Psi_H$ and $\Phi_A$ respectively. Note that $\hat\Psi$ contains a second-order time-derivative. In the first four sections of this paper we will frequently use a linear combination of these variables that remains first-order in time derivatives of perturbations, due to a cancellation between the $\dot V$ terms:
\begin{equation}
\hat\Gamma = \frac{1}{k}\left(\dot{\hat\Phi}+{\cal H}\hat\Psi\right) \label{gamma_def}
\end{equation}
From \textsection\ref{section:extra_fields} onwards we specialise to the conformal Newtonian gauge, and hence revert to the familiar potentials $\Phi$ and $\Psi$. We introduce a shorthand notation for the components of $\delta G_{\mu\nu}$ in exact analogy to that introduced for $\delta U_{\mu\nu}$, i.e. $E_{\Delta}=-a^2 \delta G_0^0$ etc. Hereafter the left-hand sides of the perturbed Einstein equations will be denoted by:
\begin{eqnarray}
E_{\Delta}&=&2(\vec{\nabla}^2+3K)\hat\Phi-6{\cal H}k\hat\Gamma-\frac{3}{2}{\cal H}EV  \nonumber \\
E_{\Theta}&=&2\,k\hat\Gamma+\frac{1}{2}EV \nonumber\\
E_P&=&6k\frac{d\,\hat\Gamma}{d\tau}+12{\cal H}k\hat\Gamma-2(\vec{\nabla}^2+3K)(\hat\Phi-\hat\Psi)\nonumber \\
&&-3E\hat\Psi+\frac{3}{2}\left(\dot E_R-2{\cal H}E_R\right)V\nonumber\\
E_{\Sigma}&=&\hat\Phi-\hat\Psi\label{es}
\end{eqnarray}
In terms of these variables the perturbed Einstein equations are \cite{Skordis}:
\begin{eqnarray}
E_{\Delta} &=& 8\pi G a^2\sum_i\rho_i\delta_i+U_{\Delta} \label{einstein1} \\
E_{\Theta} &=& 8\pi G a^2 \sum_i(\rho_i+P_i)\theta_i +U_{\Theta} \label{einstein2}\\
E_P&=&24\pi G a^2\sum_i\rho_i\Pi_i+U_{P}\label{einstein3}\\
E_{\Sigma}&=&8\pi G a^2\sum_i (\rho_i+P_i)\Sigma_i+U_{\Sigma}\label{einstein4}
\end{eqnarray}
For simplicity we will hereafter consider only the case of a universe with zero spatial curvature, $K=0$. 

\section{The General Parameterization - No Extra Fields}
\label{section:no_extra_fields}
\subsection{General case - unmodified background} 
\label{subsection:general_no_extra_fields}
Let us begin with the simplest case by applying two restrictions: i) We consider the case of modifications to gravity that appear only at the perturbative level, that is, they maintain the background equations of GR for a Friedmann-Robertson-Walker metric; ii) there are no new d.o.f. present in the theory, so  $\delta U_{\mu\nu}$ contains only metric perturbations.
We will relax restriction i) in \textsection\ref{subsection:no_extra_fields_mod_background} and restriction ii) in \textsection\ref{section:extra_fields}.
We will see shortly that the treatment presented in this subsection is also applicable to $\Lambda$CDM, because the $X+Y$ terms in eqns.(\ref{Bianchi1}) and (\ref{Bianchi2}) vanish for a cosmological constant. The agreement between an exact $\Lambda$CDM background and current data means that theories obeying restrictions i) and ii) are of particular interest, even though they correspond to a limited region of theory space.

The requirement of gauge form-invariance places strong restrictions on the forms that $\delta U_{\mu\nu}$ can take \cite{Skordis}. We will postpone a detailed discussion of these restrictions until \textsection\ref{subsection:no_extra_fields_mod_background}, where they will be a useful tool in guiding us to allowed combinations of metric perturbations. In this subsection  it suffices to point out that the standard Einstein field equations of GR are of course already gauge form-invariant; so any additive modification like $\delta U_{\mu\nu}$ must be independently gauge form-invariant in order to preserve the invariance of the whole expression. This property is a direct consequence of the fact that we have not yet modified the background equations. In this case the only objects that can be present in the tensor $\delta U_{\mu\nu}$ are the gauge-invariant metric potentials $\hat\Phi$ and $\hat\Gamma$. 

So, we can construct the tensor  $\delta U_{\mu\nu}$ from series of all the possible derivatives of $\hat\Phi$ and $\hat\Gamma$. This structure should be general enough to encompass any metric theories, where the action is constructed purely from curvature invariants, e.g. $f(R)$ gravity, Gauss-Bonnet gravity \cite{Nojiri_Gauss_Bonnet}  and Lovelock gravity \cite{Charmousis}. 
If we wish to parameterize only second-order theories then we will need to truncate these series at $N=2$, as discussed in section \ref{section:formalism}. The components of $U_{\mu\nu}$ are given by:
 \begin{eqnarray}
U_{\Delta} = \sum^{N-2}_{n=0} k^{2-n} \left(A_n \hat\Phi^{(n)} + F_n \hat\Gamma^{(n)}\right) \label{Udelta}\\
U_{\Theta} =  \sum^{N-2}_{n=0}  k^{1-n} \left(B_n \hat\Phi^{(n)} + I_n  \hat\Gamma^{(n)}\right) \label{Utheta} \\
U_{P} = \sum^{N-1}_{n=0} k^{2-n} \left(C_n \hat\Phi^{(n)}  +  J_n \hat\Gamma^{(n)}\right) \label{Up}\\
U_{\Sigma} =  \sum^{N-1}_{n=0}  k^{-n} \left(D_n \hat\Phi^{(n)} + K_n \hat\Gamma^{(n)}\right) \label{Usigma}
\end{eqnarray}
The coefficients $A_n$-$K_n$ are functions of the scale factor $a$, wavenumber $k$ and background quantities such as $\dot\rho_i$ -- for the sake of clarity we will suppress these dependencies throughout. The factors of $k$ ensure that the coefficient functions are dimensionless.

Let us take a moment to explain the upper limits on the summations in eqns.(\ref{Udelta})-(\ref{Usigma}). $\hat\Phi$ and $\hat\Gamma$ are first-order in time derivatives (see eqns.(\ref{phi_def}) and (\ref{gamma_def})). $U_{\Delta}$ is differentiated in eqn.(\ref{Bianchi1}), so truncating the series in eqn.(\ref{Udelta}) at $\hat\Phi^{(N-2)}$ gives field equations containing time derivatives of order N. $U_{\Theta}$ is treated analogously to $U_{\Delta}$. As $U_P$ and $U_{\Sigma}$ are not differentiated in the components of the Bianchi identity, the series in eqns.(\ref{Up}) and (\ref{Usigma}) are allowed to extend one order higher than those in eqns.(\ref{Udelta}) and (\ref{Utheta}).

We substitute our forms for $U_{\Delta}$, $U_{\Theta}$, $U_{P}$ and $U_{\Sigma}$ into the components of the Bianchi identity (\ref{Bianchi1}) and (\ref{Bianchi2}). $\hat\Phi$ and $\hat\Gamma$ are non-dynamical fields and so will not evolve in the absence of source terms.
Yet when we perform the substitution, the Bianchi identity appears to give us two evolution equations for $\hat\Phi$ and $\hat\Gamma$. The only way this can be avoided is if the coefficients of each term $\hat\Phi^{(n)}$ and $\hat\Gamma^{(n)}$ vanish individually, which provides us with constraint equations on the functions  $A_n$-$K_n$ (this procedure will be clarified with an example shortly). Each component of the Bianchi identity results in N constraint equations from each of the $\hat\Phi^{(n)}$ terms and $\hat\Gamma^{(n)}$ terms, and eqns.(\ref{Udelta})-(\ref{Usigma}) contain 8N-4 coefficient functions in total. Hence we have $4\textrm{N}-4$ free functions with which to describe the theory. 

\subsection{Second-order case - unmodified background}
\label{subsection:no_fields_2nd_order}
In Appendix \ref{app:constraints} we give formulae for generating the constraint equations of an arbitrary-order theory with unmodified background equations. We will now explicitly present the second-order case, which corresponds to setting N=2 in eqns.(\ref{Udelta})-(\ref{Usigma}). In a general case this will give us four free functions. However, if the background equations are unaltered then we must set \mbox{$F_0=I_0=0$} because $\hat\Gamma$ contains a second-order conformal time derivative of the scale factor. One might consider cancelling this $\ddot a$ term by adding a term proportional to $EV$, but this would break the gauge-invariance of the perturbed Einstein equations. We will see later that modification of the background equations allows us to add an $EV$ term without violating gauge-invariance, which in turns means that $\hat\Gamma$ can be present in $U_{\Delta}$ and $U_{\Theta}$. 
 
Using eqns.(\ref{Bianchi1}) and (\ref{Bianchi2}) we find that setting $F_0=I_0=0$ forces $J_1=K_1=0$ also. Then, for the second-order case, the remaining terms in $\delta U_{\mu\nu}$ are:
 \begin{eqnarray}
 \label{2nd_order_Us}
U_{\Delta} &=& A_0 k^2\hat\Phi\nonumber\\
U_{\Theta} &=& B_0 k\hat\Phi\nonumber \\
U_{P} &=& C_0 k^2\hat\Phi + C_1 k\dot{\hat\Phi} +J_0 k^2\hat\Gamma \nonumber\\
U_{\Sigma} &=&  D_0 \hat\Phi+\frac{D_1}{k} \dot{\hat\Phi} + K_0 \hat\Gamma \label{U_2nd_order}
\end{eqnarray}
The constraint equations are given in Table \ref{tab:2nd_order_constraints}, indicating the terms and Bianchi identity from which they arise ($B1\Rightarrow$ eqn.(\ref{Bianchi1}), $B2\Rightarrow$ eqn.(\ref{Bianchi2})). These expressions can be generated using the formulae in Appendix \ref{app:constraints}. We can see immediately that the $\hat\Gamma$ terms in $U_P$ and $U_{\Sigma}$ vanish, leaving $\delta U_{\mu\nu}$ expressed entirely in terms of $\hat\Phi$ and $\dot{\hat\Phi}$. We have two free functions remaining, which we will choose to be $D_0$ and $D_1$. 
Eliminating $C_1$ from the two $\dot{\hat\Phi}$ constraints gives (where ${\cal H}_k={\cal H}/k$):
\begin{equation}
\frac{\cal H}{k}D_1=-\frac{1}{2} \left(A_0+3 {\cal H}_k B_0\right) 
\end{equation}
The combination on the right-hand side appears when we form the (Fourier-space) Poisson equation from eqns.(\ref{einstein1}) and (\ref{einstein2}), where it acts to modify the value of Newton's gravitational constant:
\begin{equation}
-k^2\hat\Phi = 4\pi \frac{G_0}{1-\tilde g} a^2 \sum_i \rho_i \Delta_i 
\label{Poisson} 
\end{equation}
where $\Delta_i=\delta_i + 3{\cal H}(1+w_i)\theta_i $ is a gauge-invariant matter perturbation and 
\begin{equation}
\label{gtilde}
\tilde{g}=-\frac{1}{2}\left(A_0+3 {\cal H}_k B_0\right)
\end{equation}
 The sum in eqn.(\ref{Poisson}) is over all known fluids and dark matter, and $G_0$ denotes the canonical value of Newton's constant. 
 From here on we will replace $D_1/k$ in eqn.(\ref{2nd_order_Us}) by $\tilde{g}/{\cal H}$ to remind us of the connection between the modifications to the slip relation and the Poisson equation. We will also replace $D_0$ by $\zeta$ to distinguish it from the other coefficient functions, which can all be expressed in terms of $\tilde{g}$ and $\zeta$ using the constraint equations. We continue to suppress the arguments of $\tilde g$ and $\zeta$.
 \begin{table}
\begin{tabular}{| c | c | l |}
\hline
$\qquad$  & Origin & \qquad\; Constraint equation  \\ \hline
1 & \;[B1] $\;\hat\Phi$ \,& $\dot A_0+{\cal H} A_0+k B_0+{\cal H} C_0 = 0$\\ \hline
2 & \;[B1] $\;\dot{\hat\Phi}$ \,& $A_0+{\cal H}_k C_1 = 0 $\\ \hline
3 & \;[B1] $\;\hat\Gamma$ \,& $J_0=0 $\\ \hline
4 & \;[B2] $\;\hat\Phi$ \,& $\dot B_0+2{\cal H} B_0-\frac{1}{3}k C_0+\frac{2}{3}k D_0 = 0$\\ \hline
5 & \;[B2] $\;\dot{\hat\Phi}$ \,& $B_0-\frac{1}{3} C_1+\frac{2}{3} D_1= 0$\\ \hline
6 & \;[B2] $\;\hat\Gamma$ \,& $2K_0-J_0=0 $\\ \hline
\end{tabular}
\caption{Table of the constraint equations for the second-order metric theory specified in \textsection\ref{subsection:no_fields_2nd_order}. These can be generated using the formulae in Appendix \ref{app:constraints}.}
\label{tab:2nd_order_constraints}
\end{table}
 
The effective gravitational constant appearing in the Poisson equation is \mbox{$G_{\mathrm{eff}}=G_0/(1-\tilde g)$}.
The traceless space-space component of the Einstein equations becomes: 
\begin{equation}
\label{2nd_order_slip}
\hat\Phi-\hat\Psi=8\pi G_0 \sum_i (\rho_i+P_i)\Sigma_i +\zeta\hat\Phi+\frac{\tilde g}{\cal H}\dot{\hat\Phi}
\end{equation}
The anisotropic stress perturbation $\Sigma_i$ is automatically gauge-invariant, but negligible for standard fluids at late times. The above expression echoes its PPN equivalent, eqn.(\ref{PPNslip}); but note that, as discussed in \textsection\ref{section:conventional}, $\zeta$ is a function of background quantities (which potentially introduce time- and scale-dependence), whereas $\zeta_{\tt{PPN}}$ depended only upon fundamental parameters of a gravitational theory.

Other authors have made numerous different choices for the two free functions of a second-order theory; a useful summary of some of these is provided by \cite{Daniel2010} . A common choice is to introduce a function $Q=G_{\mathrm{eff}}/G_0$, related to our $\tilde g$ by $Q=(1-\tilde g)^{-1}$ \cite{Bean} (though different notation is in no short supply) . The relationship between the two potentials is often parameterized as \mbox{$\hat\Phi=\eta_{\mathrm{slip}}(a,k)\,\hat\Psi$} 
in the spirit of the PPN parameter $\gamma_{PPN}$ \cite{Will1971}. It might be felt that by introducing yet another parameterization of PPF we are adding to this disarray. However, in the next subsection we will argue that a two-function slip relation such as eqn.(\ref{2nd_order_slip}) is needed to avoid implicitly introducing higher-order derivatives into a purely metric theory.

Writing the relationship between the two gauge-invariant potentials as \mbox{$\hat\Phi=\eta_{\mathrm{slip}}(a,k)\hat\Psi$} implies that the spatial off-diagonal component of the Einstein field equations is: 
\begin{equation}
\hat\Phi-\hat\Psi = \left(1-\frac{1}{\eta_{\mathrm{slip}}}\right)\hat\Phi
\label{eta_slip}
\end{equation}
Comparing the above equation with eqn.(\ref{2nd_order_slip}) implies:
\begin{equation}
\eta_{\mathrm{slip}}^{-1}=1-\zeta-\frac{\tilde g}{\cal H} \frac{d\,\mathrm{ln}\hat\Phi}{d\tau}
\end{equation}
Now $\eta_{\mathrm{slip}}$ has an environmental dependence, which is problematic. We would require detailed knowledge of the environment in which we wish to test a theory  \textit{a priori}, and the PPF functions would need to be recalculated for numerous different situations. Unless $\dot{\hat\Phi}=0$, the parameterizations in eqns.(\ref{2nd_order_slip}) and (\ref{eta_slip}) do not have a simple equivalence.

A degeneracy arises between $\tilde g$ and $\zeta$ when comparing to data from weak gravitational lensing, which probes the combination $\Phi+\Psi$ in the conformal Newtonian gauge. In parameterizations equivalent to \mbox{$(Q,\eta_{\mathrm{slip}})$} the degeneracy is \mbox{$Q(1+1/\eta_{\mathrm{slip}})$}, so for lensing applications it makes sense to define new parameters along and perpendicular to the degeneracy direction \cite{Song2010, Zhao2010, DanielLinder}.
 In the $(\tilde g,\zeta)$ parameterization a degeneracy remains. The dominant contributions to lensing signals come from quasistatic scales, on which time derivatives of perturbations can be neglected (see later for a fuller discussion). The degeneracy is then:
\begin{equation}
-k^2 (\Phi+\Psi) = 4\pi G_0 a^2 \sum_i\rho_i\delta_i\frac{\left(2-\zeta\right)}{1-\tilde g}
 \label{lensing_degeneracy}
\end{equation}
It seems that neither of the two parameterizations presented so far are optimal for weak lensing constraints.

\subsection{Why neglecting $\tilde g$ in the slip relation implies a higher-derivative theory}
\label{subsection:why_gtilde}
We have seen in the previous section that in a metric-based second-order theory of modified gravity the most general form of the gravitational slip should be expressed in terms of the gauge-invariant potential $\hat\Phi$ \textit{and} its first derivative with respect to conformal time. Two free functions $\zeta$ and $\tilde g$ were used as the coefficients of these terms respectively, where $\tilde g$ resulted in a modification to Newton's gravitational constant in the Poisson equation. Using a single function to relate $\hat\Phi$ and $\hat\Psi$ is equivalent to setting $\tilde g=0$ (see eqn.(\ref{2nd_order_slip})), which is inconsistent with with allowing a second free function to modify Newton's constant. Making the choice $\tilde g=0$ uses up one degree of freedom, leaving us only a single free function with which to describe the system.

The above reasoning is set within the confines of a second-order theory. We will now show that using a single function to relate $\hat\Phi$ and $\hat\Psi$ whilst maintaining \mbox{$G_{\mathrm{eff}} \neq G_0$} is equivalent to invoking a higher-derivative theory of gravity. To do this, let us consider the form that the tensor $\delta U_{\mu\nu}$ would take in a third-order theory. Its components would be:
\begin{eqnarray}
U_{\Delta} &=& A_0 k^2 \hat\Phi+A_1 k \dot{\hat\Phi}+F_0 k^2\hat\Gamma\nonumber\\
U_{\Theta} &=& B_0 k \hat\Phi+B_1 \dot{\hat\Phi} + I_0 k\hat\Gamma\nonumber \\
U_{P} &=& C_0 k^2 \hat\Phi+C_ 1 k \dot{\hat\Phi}+C_2 \ddot{\hat\Phi}+J_0 k^2\hat\Gamma+J_1k\dot{\hat\Gamma} \nonumber \\
U_{\Delta} &=& D_0 \hat\Phi +\frac{D_1}{k}\hat{\dot\Phi}+\frac{D_2}{k^2}\ddot{\hat\Phi}+K_0 \hat\Gamma+\frac{K_1}{k}\dot{\hat\Gamma}  \label{3rd_order_Us} 
\end{eqnarray}
The constraint equations for this system are given in \mbox{Table~\ref{tab:3rd_order_constraints}}. 
We will continue to define the combination that modifies $G_0$ as $\tilde g= -0.5(A_0+3{\cal H}_k B_0)$, but note that this is no longer equal to ${\cal H}_k D_1$ as it was in the second-order case.
\begin{table}
\begin{tabular}{| c | c | l |}
\hline
 $\qquad$ & Origin & \qquad\qquad Constraint equation  \\ \hline
1 & \,[B1] $\;\hat\Phi$ \,&$\dot A_0+{\cal H}A_0+k B_0+{\cal H}C_0=0$ \\ \hline
2 & \,[B1] $\;\dot{\hat\Phi}$ \,&$\dot A_1+{\cal H}A_1+k A_0+k B_1+{\cal H}C_1=0$ \\ \hline
3 & \,[B1] $\;\ddot{\hat\Phi}$ \,&$k A_1+{\cal H}C_2=0$ \\ \hline
4 & \,[B1] $\;\hat\Gamma$ \,&$\dot F_0+{\cal H}F_0+k I_0+{\cal H} J_0=0$ \\ \hline
5 & \,[B1] $\;\dot{\hat\Gamma}$ \,&$kF_0+{\cal H} J_1=0$ \\ \hline
6 & \,[B2] $\;\hat\Phi$ \,&$\dot B_0+2{\cal H}B_0-\frac{1}{3}kC_0+\frac{2}{3}k D_0=0$ \\ \hline
7 & \,[B2] $\;\dot{\hat\Phi}$ \,&$\dot B_1+k B_0+2{\cal H}B_1-\frac{1}{3}k C_1+\frac{2}{3}k D_1=0$ \\ \hline
8 & \,[B2] $\;\ddot{\hat\Phi}$ \,&$ B_1-\frac{1}{3}C_2+\frac{2}{3} D_2=0$ \\ \hline
9 & \,[B2] $\;\hat\Gamma$ \,&$\dot I_0+2{\cal H} I_0 -\frac{1}{3}k J_0+\frac{2}{3}k K_0=0$ \\ \hline
10 & \,[B2] $\;\dot{\hat\Gamma}$ \,&$I_0-\frac{1}{3} J_1+\frac{2}{3} K_1=0$ \\ \hline
\end{tabular}
\caption{Table of the constraint equations for the third-order metric theory specified in \textsection\ref{subsection:why_gtilde}. These can be generated using the formulae in Appendix \ref{app:constraints}.}
\label{tab:3rd_order_constraints}
\end{table}
Consider the case where we set $D_1=D_2=K_0=K_1=0$, that is, we use a single function to relate $\hat\Phi$ and $\hat\Psi$. Through linear combinations of the constraints in \mbox{Table \ref{tab:3rd_order_constraints}} we derive the expressions:
\begin{eqnarray}
&&A_1+3{\cal H}_k B_1=0 \label{A1B1}\\
&&F_0+3{\cal H}_k I_0=0 \label{E0F0}\\
&&\tilde g = \frac{3}{2} B_1 \left({\cal H}_k^2+\frac{1}{3}-\frac{\dot{\cal H}_k}{k}\right) =\frac{1}{2k^2} B_1\left(\frac{3 E}{2}+k^2\right) \label{gtilde_B1} 
\end{eqnarray}
The first two of these expressions are the combinations that appear when we form the Poisson equation. They indicate that the potential additive modifications proportional to $\dot{\hat\Phi}$ and $\hat\Gamma$ disappear; the format of eqn.(\ref{Poisson}) is retained. Eqn.(\ref{gtilde_B1}) shows that we can only have a modification to the effective gravitational constant if $B_1\neq 0$, and so from eqn.(\ref{A1B1}) $A_1 \neq 0$ also. Using the third equation in Table \ref{tab:3rd_order_constraints}, $C_2 \neq 0$ in this case. Hence we are forced to include $\dot{\hat\Phi}$ terms in $U_{\Delta}$ and $U_{\Theta}$, and a $\ddot{\hat\Phi}$ term in $U_P$. Since $\hat\Phi$ contains a first-order time derivative already (see eqn.(\ref{phi_def}), the $\dot U_{\Delta}$ in eqn.(\ref{Bianchi1}) will result in field equations containing third-order time derivatives - a higher-order gravitational theory.

This result is a direct consequence of choosing $D_1=0$ in $U_{\Sigma}$. Removing this constraint changes eqn.(\ref{gtilde_B1}) to:
\begin{equation}
\tilde g = \frac{1}{2 k^2} B_1 \left(\frac{3 E}{2}+k^2\right) +{\cal H}_k D_1
\label{gtilde_B1_D1} 
\end{equation}
which permits $B_1=0, \tilde g \neq 0$, as we had in  \textsection{\ref{subsection:no_fields_2nd_order}}.

The above findings make sense within the context of the Lovelock-Grigore theorem \cite{Lovelock1, Lovelock2}, which states that under the assumptions of four-dimensional Riemannian geometry and no additional fields, the Einstein-Hilbert action (plus a cosmological constant) is the only possible action that leads to local second-order field equations. In eqn.(\ref{2nd_order_Us}) the presence of $D_1/k$ in $U_{\Sigma}$ means that this parameterization implies a non-local theory. This is not in itself problematic -- nonlocal theories can arise when a degree of freedom has been integrated out, or eliminated from the action using an integral solution of the corresponding equation of motion.
If \mbox{$D_1=D_2=K_1=0$} in eqns.(\ref{3rd_order_Us}) there are no nonlocal terms present in the gravitational field equations, so we should not be surprised that the Lovelock-Grigore theorem prevents us from obtaining a second-order theory. In \textsection\ref{section:examples} we will meet theories which evade the Lovelock-Grigore theorem in a number of different ways: by introducing new d.o.f. (e.g. scalar-tensor theory), higher-order field equations \big($f(R)$ gravity\big), or through nonlocality and extra dimensions (DGP).

\subsection{Cases with `XY' backgrounds}
\label{subsection:no_extra_fields_mod_background}
The previous examples have all assumed that the background field equations are those of a Friedmann-Robertson-Walker metric plus standard cosmological fluids. We now relax this assumption and consider theories which modify the Einstein field equations at both the background and perturbative levels. It is well-known that any modification to background-level field equations is indistinguishable from the effects of a dark fluid \cite{Hu_Eisenstein_1999}; hence we can write any background equations as the standard FRW ones with an additional energy density and pressure, see eqns.(\ref{E_F}) and (\ref{E_R}). We will refer to such theories as having `XY backgrounds'.

Any extension to GR must preserve the property of diffeomorphism invariance. Invariance under passive diffeomorphisms corresponds to the familiar principle of general covariance. Applying a passive diffeomorphism will generally result in field equations which look different to those in the old co-ordinate system. In contrast, invariance under \textit{active} diffeomorphisms requires that the actual form of field equations remains unchanged by a gauge transformation. In Table \ref{tab:gauge_transf} we list the gauge transformations for relevant variables. In practical terms, gauge form-invariance means that the extra terms that appear under a gauge transformation must cancel each other (using identities from the zero-order field equations if need be). This places tight restrictions on our form for $\delta U_{\mu\nu}$.

To see how this happens in ordinary GR, consider the linearly perturbed `00' component of the Einstein equations (that is, eqn.(\ref{einstein1}) with $U_{\Delta}$ set to zero). When we apply a gauge transformation the left-hand side acquires a term $-3{\cal H}E \xi/a$. This is cancelled by the transformation of $\delta$ on the right-hand side, provided that \mbox{$E=E_F+E_R=8\pi G a^2 \sum_i\rho_i(1+w_i)$}, i.e. provided that the zeroth-order equations are satisfied. This is why we were only able to use gauge-invariant potentials in $\delta U_{\mu\nu}$ in \textsection\ref{subsection:general_no_extra_fields} and \textsection\ref{subsection:no_fields_2nd_order}:  if we don't alter the zeroth-order equations, adding anything else breaks gauge form-invariance.
\begin{table}
\begin{tabular}{| c | c |}\hline
  Metric variables & Fluid variables  \\ \hline
$\Xi \rightarrow \Xi-\frac{\dot\xi}{a}$ & $\delta \rightarrow \delta -\frac{3}{a}(1+w){\cal H}\xi$ \\ \hline
$\epsilon\rightarrow\epsilon+\frac{1}{a}[\xi+{\cal H}\psi-\dot\psi]$ & $\theta \rightarrow \theta+\frac{1}{a}\xi$ \\ \hline
$\beta \rightarrow \beta +\frac{1}{a}[6{\cal H}\xi-2k^2\psi]$ & $\Pi \ \rightarrow \Pi +\frac{1}{a} [\dot w-3w{\cal H}(1+w)]\xi$ \\ \hline
$\nu \rightarrow \nu+\frac{2}{a}\psi$ & $ \Sigma \rightarrow \Sigma$ \\ \hline  
$V \rightarrow V+\frac{2}{a}\xi$ & $$ \\ \hline
\multicolumn{2}{|c|}{}\\ \hline
\multicolumn{2}{|c|}{Components of $\delta G_{\mu\nu}$} \\ \hline
$E_{\Delta} \rightarrow E_{\Delta}-\frac{3}{a}{\cal H}\left(E_F+E_R\right)\xi$ & $E_{\Theta} \rightarrow E_{\Theta}+\frac{1}{a}\left(E_F+E_R\right)\xi$ \\ \hline
${E_{P} \rightarrow E_{P}+\frac{3\xi}{a}(\dot E_R-2{\cal H}E_R)}$ & $E_{\Sigma} \rightarrow E_{\Sigma}$ \\ \hline
\end{tabular}
\caption{Behaviour of metric and fluid variables under infinitesimal diffeomorphisms generated by the vector field $\xi_{\mu}=a (-\xi,\vec{\nabla}_i\psi)$. Note that the shear $\Sigma$ is gauge-invariant.} 
\label{tab:gauge_transf}   
\end{table}

Now that we wish to consider XY backgrounds this procedure no longer works, because $E\neq 8\pi G_0 a^2\sum_i \rho_i(1+w_i)$. We must add a new term to $U_{\Delta}$ that will produce a part like $\frac{3}{a}{\cal H}a^2(X+Y)\xi$ under a gauge transformation. Only then will the gauge-variant parts cancel by virtue of the zeroth-order equation.

As a toy example, consider a simple theory which modifies the zeroth-order equations solely by introducing time-dependence to Newton's gravitational constant. Following the notation of previous sections, we can write the sum of the modified Friedmann and Raychaudhuri equations as:
\begin{equation}
\label{toy_example_background}
E=8\pi\frac{G_0}{1-{\tilde g_b}(a)} a^2\sum_i\rho_i(1+w_i)
\end{equation}
Rewriting this in the form of ordinary GR (and hereafter suppressing the argument of $\tilde g_b$):
\begin{equation}
\label{toy_example_perturbed}
E=8\pi G_0 a^2\sum_i\rho_i(1+w_i)+\tilde g_b E
\end{equation}
from which we can identify $a^2(X+Y)=\tilde g_b E$ (see eqns.(\ref{E_F}) and (\ref{E_R})). A possible form for $U_{\Delta}$ is then:
\begin{equation}
\label{U_Delta_toy_example}
U_{\Delta}=-6\tilde g_b\,{\cal H}\left[k\hat\Gamma+\frac{1}{4}E V\right]+A_0k^2\hat\Phi
\end{equation}
Unlike the second-order example of \textsection\ref{subsection:no_fields_2nd_order}, $U_{\Delta}$ now contains a $\hat\Gamma$ term.
The offending second-order derivative of the scale factor present in $\hat\Gamma$ is cancelled by the term proportional to $EV$. We did not have the freedom to add such a term in the case of unmodified background equations, because $V$ is gauge-variant.

Using eqn.(\ref{U_Delta_toy_example}) in eqn.(\ref{einstein1}) and the transformations given in \mbox{Table \ref{tab:gauge_transf}}, it can be verified that the gauge-variant parts cancel by satisfying eqn.(\ref{toy_example_background}). Note that the need to have gauge-invariant second-order equations has totally fixed the first term in eqn.(\ref{U_Delta_toy_example}); all freedom resides in the gauge-invariant part of $U_{\Delta}$ via the function $A_0$. 

As a quick sanity check, one can verify that the field equations remain second order in the conformal Newtonian gauge. In this gauge $V=\dot V=0$, $\Xi=-\Psi$ and $\beta=-6\Phi$, so $\hat\Phi$ and $\hat\Psi$ reduce to their familiar counterparts $\Phi$ and $\Psi$ that appear in the linearly perturbed FRW metric (recall that \mbox{$\hat\Gamma=1/k (\dot{\hat\Phi}+{\cal H}\Psi)$}). $U_{\Delta}$ is then explicitly first-order in derivatives, resulting in second-order equations. 

$U_{\Theta}$ is treated analogously to $U_{\Delta}$, and has the form:
\begin{equation}
\label{U_theta_mod_background}
U_{\Theta}=2\tilde g_b\,\left[k\hat\Gamma+\frac{1}{4}E V\right]+B_0k\hat\Phi
\end{equation}
Combining eqns.(\ref{U_Delta_toy_example}) and (\ref{U_theta_mod_background}),  we find that the Poisson equation has the same form as it did in the case with unmodified background equations, eqn.(\ref{Poisson}). We define the combination \mbox{$\tilde g =-1/2\left(A_0+3{\cal H}_kB_0\right)$} as we did in \textsection{\ref{subsection:no_fields_2nd_order}}. We will \textit{assume} that the modifications to Newton's gravitational constant appearing in the zeroth-order and perturbed equations are the same, i.e. \mbox{$\tilde g_b=\tilde g$}, noting that we have not formally proved this to be the case.

Once we have deduced the form of $U_{\Delta}$ and $U_{\Theta}$, $U_P$ and $U_{\Sigma}$ can be found using the Bianchi identities \big(eqns.(\ref{Bianchi1}) and (\ref{Bianchi2})\big). Eliminating the free function $A_0$ in favour of $\tilde g$ and $B_0$, these are:
\begin{eqnarray}
{\cal H} U_P &=&3\hat\Phi\left[k{\cal H}\dot B_0+kB_0 L+\frac{2}{3} k^2\left(\dot{\tilde g}+{\cal H}\tilde g\right)\right]\nonumber\\
&&+3\dot{\hat\Phi}\left[k{\cal H}B_0+\tilde g \left(\frac{E}{2} +\frac{k^2}{3}\right)\right]\nonumber\\
&&+6\left(\dot{\tilde g}{\cal H}+\tilde g L\right)\left[k\hat\Gamma+\frac{1}{4}EV\right] \nonumber\\
&&+6\tilde g {\cal H}\frac{d}{d\tau}\left[k\hat\Gamma+\frac{1}{4}EV\right] \nonumber \\
&&\frac{3}{2}\tilde g E V \left[\frac{k^2}{3}-\dot{\cal H}\right]-\frac{3}{2}E\tilde g {\cal H}\dot V
\end{eqnarray}
\begin{eqnarray}
U_{\Sigma}&=&\frac{\hat\Phi}{2{\cal H}k}\left[2k\left(\dot {\tilde g}+{\cal H}\tilde g\right)-3B_0\left(\frac{E}{2}+\frac{k^2}{3}\right)\right]\nonumber\\
&&-\frac{\tilde g}{\cal H}\left(k\hat\Gamma-\dot{\hat\Phi}\right)\label{Usigma_toy_example}\nonumber\\
&&\hspace{-1cm}\mathrm{where}\quad L={\cal H}^2+\dot{\cal H}-\frac{k^2}{3}.
\end{eqnarray}
As expected, we find that $U_{\Sigma}$ contains only gauge-invariant perturbation variables. This must be the case since all other terms in eqn.(\ref{einstein4}) are gauge-invariant, so there is nothing to cancel against. 

If we express the last term in $U_{\Sigma}$ in terms of $\hat\Psi$ we find that we \textit{can} write the relationship between the two potentials as $\hat\Phi=\eta_{\mathrm{slip}}(a,k)\hat\Psi$ for this toy example. We have already chosen our two PPF functions to be $\tilde g$ and $B_0$, so $\eta_{\mathrm{slip}}$ is simply a particular combination of these:
\begin{equation}
\label{toy_example_slip}
\eta_{\mathrm{slip}}=\frac{{\cal H}(1-\tilde g)}{{\cal H}(1-\tilde g)-\dot{\tilde g}+\frac{3 B_0}{2 k}\left(\frac{E}{2}+\frac{k^2}{3}\right)}
\end{equation}

The key result of this toy example is that (for purely metric theories) $\delta U_{\mu\nu}$ can have a more complex form when the background equations are not standard FRW plus standard cosmological components (baryons, CDM, etc.) The hierarchy of constraint equations becomes more complex due to the non-zero $X$ and $Y$ terms in eqns.(\ref{Bianchi1}) and (\ref{Bianchi2}). We can use the principles of energy conservation, gauge-invariance and second-order field equations as a shortcut to the correct forms; the same results would be obtained by solving the hierarchy of constraint equations directly.%

\section{The General Parameterization- extra fields}
\label{section:extra_fields}
The formalism we have developed so far is only applicable to purely metric theories. Yet, as discussed in \textsection\ref{section:conventional}, the majority of modified gravity theories introduce new degrees of freedom, often as additional scalar, vector or tensor fields. It is not immediately obvious that the behaviour of these theories can be be encapsulated by a either a $(Q,\eta_{\mathrm{slip}})$ or $(\tilde{g},\zeta)$ parameterization. There is a risk that we might develop a model-independent formalism that does not map onto most of our well-studied theories. 

In \textsection\ref{section:examples} we will study several example cases, chosen to be representative of common classes of modified gravity models, and ask whether they can be expressed in the two-function format of quasistatic PPF. Attempting to map disparate theories onto a single framework is only plausible if those theories share some common features. Hence, before turning to specific examples, we wish to consider what general statements can be made about the structure of the the field equations in theories with extra degrees of freedom.

Let the scalar perturbations to an extra degree of freedom be denoted by $\chi$, e.g. if the new degree of freedom is a scalar field $\phi$ then $\chi=\delta\phi$. (We have succumbed to the common but unwise choice of terminology by referring to `scalar perturbations', even though the new degree of freedom may itself be a vector or tensor field. `Spin-0 perturbations' would be a better choice of terminology \cite{Lim_Lorentz_CMB}).  If we are to obtain second-order field equations then we know that $U_{\Delta}$ and $U_{\Theta}$ can only contain the perturbations $\chi$ and $\dot\chi$. 

More generally we can introduce multiple new d.o.f. and denote their scalar perturbations by the vector $\vec{\chi}$, with components $\chi^{(i)}$. 
The perturbed field equations in a general gauge are awkward and rarely used; hence we will specialise to the conformal Newtonian gauge for the remainder of this paper. The relevant expressions for scalar-tensor theory (\textsection\ref{subsection:scalar_tensor}) are presented in a general gauge in Appendix \ref{app:gauge_invariance}.

In the conformal Newtonian gauge the 00-element of the tensor $\delta U_{\mu\nu}$ can be represented as: 
\begin{eqnarray}
\label{Udelta_general}
U_{\Delta} &=& k^2 \vec{\alpha}_0^{\mathbf{T}}\vec{\chi}+k \vec{\alpha}_1^{\mathbf{T}}\dot{\vec{\chi}}+A_0 k^2 \Phi+A_1 k\,\dot\Phi\nonumber\\
&&+F_0 k^2\Psi+F_1 k\dot\Psi+\frac{M_0}{k^2}\delta\rho+\frac{M_1}{k^3}\delta\dot\rho
\end{eqnarray}
where $\delta\rho$ is the total energy density fluctuation of standard cosmological fluids (similarly for $\theta, \,\Pi$ and $\Sigma$ to be used shortly).
$\vec{\alpha}_0$ and $\vec{\alpha}_1$ denote vectors of functions with the same dimensionality as $\vec{\chi}$. $\vec{\alpha}_i, A_i, F_i$ and $M_i$ are functions of background quantities such as $\rho$ and $a$; these dependencies have been suppressed for clarity. 
Note that eqn.(\ref{Udelta_general}) has the form indicated schematically in eqn.(\ref{U_decomposition}).
The term $M_1\delta\dot\rho$ represents a modification which depends on the rate of change of the density fluctuations of ordinary matter. Whilst formally this term is permitted to be present in $U_{\Delta}$, we are unaware of any theory of modified gravity that results in a perturbed 00-equation with a term like this. Theories employing a chameleon mechanism introduce modifications to GR that depend upon the environmental matter density, but not upon its $rate$ of change . We will therefore choose $M_1=0$ in what follows. The $\delta\rho$ term in eqn.(\ref{Udelta_general}) can be eliminated using the relation:
\begin{equation}
8\pi G_0 a^2\delta\rho=E_{\Delta}-U_{\Delta}
\end{equation}
$E_{\Delta}$ can be expressed in terms of $\Phi, \dot\Phi$ and $\Psi$ in the conformal Newtonian gauge, see eqns.(\ref{es}). Rearranging and redefining the coefficient functions, we obtain:
\begin{eqnarray}
\label{rescaled_general_Udelta2} 
U_{\Delta} = k^2 \vec{\alpha}_0^{\mathbf{T}}\vec{\chi}+k \vec{\alpha}_1^{\mathbf{T}}\dot{\vec{\chi}}+A_0 k^2 \Phi&+&A_1 k\,\dot\Phi\nonumber\\
&+&F_0 k^2\Psi+F_1k\dot\Psi
\end{eqnarray}
This procedure has enabled us to eliminate energy density fluctuations of ordinary matter from $U_{\Delta}$. If the new d.o.f. are not coupled to ordinary matter then $\delta\rho$ does not appear in $U_{\Delta}$ anyway. It is worth reminding the reader that in this section we are working in the conformal Newtonian gauge, so $\Phi$ and $\Psi$ should not be confused with their gauge-invariant counterparts $\hat\Phi$ and $\hat\Psi$, which already contain first- and second-order derivatives respectively. If we were to consider eqn.(\ref{rescaled_general_Udelta2}) in a general gauge we would find that $\dot{\hat\Phi}$ and $\hat\Psi$ only appear in the combination $\hat\Gamma$. The $\dot\Psi$ in eqn.(\ref{rescaled_general_Udelta2}) would become $-\dot\Xi$ \big(recall that in the conformal Newtonian gauge \mbox{$\hat\Phi \equiv \Phi \equiv -\frac{1}{6}\chi$} and \mbox{$\hat\Psi\equiv\Psi\equiv-\Xi$}, using eqns.(\ref{phi_def}) and (\ref{psi_def})\big). \newline

We apply a similar treatment to the remaining components of $\delta U_{\mu\nu}$. Recall that $U_P$ and $U_{\Sigma}$ are permitted to contain second-order terms such as $\ddot\Phi$ and $\ddot\chi$. Terms such as $\dot\theta, \dot\Pi, \ddot\Pi, \dot\Sigma$ and $\ddot\Sigma$ are discarded to maintain contain consistency with our treatment of $U_{\Delta}$; we stress again that this is done only on an intuitive basis. The resulting expressions for $\delta U_{\mu\nu}$ (together with eqn.(\ref{rescaled_general_Udelta2})) are as follows:
\begin{eqnarray}
U_{\Theta} &=&k \vec{\beta}_0^{\mathbf{T}}\vec{\chi}+\vec{\beta}_1^{\mathbf{T}}\dot{\vec{\chi}}+B_0 k\Phi+B_1\dot\Phi+I_0 k\Psi+I_1\dot\Psi\label{extrafields_theta} \\
U_{P} &=& k^2 \vec{\gamma}_0^{\mathbf{T}}\vec{\chi}+k \vec{\gamma}_1^{\mathbf{T}}\dot{\vec{\chi}}+\vec{\gamma}_2^{\mathbf{T}}\ddot{\vec{\chi}}+C_0 k^2 \Phi+C_1 k\,\dot\Phi\nonumber\\
&&+C_2\ddot\Phi+J_0 k^2\Psi+J_1k\dot\Psi+J_2\ddot\Psi\label{extrafields_P}\\
U_{\Sigma} &=&  \vec{\varepsilon}_0^{\mathbf{T}}\vec{\chi}+\frac{1}{k} \vec{\varepsilon}_1^{\mathbf{T}}\dot{\vec{\chi}}+\frac{1}{k^2}\vec{\varepsilon}_2^{\mathbf{T}}\ddot{\vec{\chi}}+D_0 \Phi+\frac{D_1}{k}\,\dot\Phi \nonumber\\
&&+\frac{D_2}{k^2}\ddot\Phi+K_0 \Psi+\frac{K_1}{k}\dot\Psi+\frac{K_2}{k^2}\ddot\Psi\label{extrafields_Sigma}
\end{eqnarray}
where $\beta_i,\ldots\varepsilon_i$ and $B_i,\ldots K_i$ denote functions of background quantities.

The Bianchi identities then give us two constraint equations coupling terms in  $\chi^{(i)}, \Phi, \Psi$ and their derivatives. In contrast to the previous section we can no longer set the coefficients of the each term to zero individually. In the case without extra fields  this was possible because all our variables were non-dynamical, so obtaining evolution equations for them would be unphysical. But now that extra fields appear in $\delta U_{\mu\nu}$, the Bianchi identities yield equations describing how the metric variables respond to the set of perturbations $\vec{\chi}$. Therefore we no longer have a hierarchy of constraint equations for the coeffcients $\alpha_0,\ldots K_2$ that allow us to reduce them down to two functions. This is not problematic in itself. To map a specific theory onto the parameterization we can simply pull the necessary coefficients out of the perturbed field equations. We will see shortly (\textsection\ref{section:examples}) that in many cases that these are relatively simple functions.

To constrain a general parameterized theory using Markov Chain Monte Carlo (MCMC) analysis we instead choose some sensible ansatz for the functions $\alpha_0,\ldots K_2$. For example, a Taylor series up to cubic order in $\Omega_{\Lambda}$ was used in \cite{Ferreira_Skordis,Zuntz2011}; the MCMC then constrains the coefficients of the terms in the Taylor series. Rigidly fixing the format of the parameterization in this way means that we simply have to constrain real numbers. This simplicity is a key advantage of explicitly parameterizing for the new fields as in eqns.(\ref{rescaled_general_Udelta2}-\ref{extrafields_Sigma}). The alternative approach -- absorbing the new fields into an evolving $G_{\mathrm{eff}}$ and slip parameter where possible -- will give $G_{\mathrm{eff}}$ and $\zeta$ very complicated forms that are difficult to parameterize (for example, see eqns.(\ref{ST_QS_Poisson}) and (\ref{fR_Poisson_QS})). The trade-off is that our method requires considerably more than two coefficient functions. We expect that some of these will be well-constrained by the data, others less so.

In the case of just one or two new d.o.f., the system consisting of the Einstein equations, the two conservation equations for ordinary matter and two Bianchi identities for the $U$-tensor can be solved. In order to avoid a contradiction, the Bianchi identities for the $U$-tensor must be equivalent to the equations of motion for the extra degrees of freedom (obtained by varying the action with respect to the extra fields or similar). Futhermore, when only a single d.o.f. is present the solutions of the two components of the Bianchi identity must be consistent with each other.

When more than two new d.o.f. are present the Bianchi identities do not provide sufficient information to solve the system, and we must supply additional relations between the new d.o.f, metric variables and matter variables. With our goal of an abstract, unified framework in mind, we will introduce a general structure to tackle such cases. 
We make the conjecture that one can use the `generalized dark matter' (GDM) formalism developed by Hu \cite{Hu_GDM} in order to obtain the necessary closure relations. 
GDM is a phenomenological model in which specification of three parameters - an equation of state, a rest-frame sound speed and a viscous sound speed - suffice to reconstruct the full perturbed stress-energy tensor of a fluid. Cold dark matter, radiation, massive neutrinos, WIMPs, scalar fields and a cosmological constant can all be recovered as limiting cases of GDM. In our case the d.o.f. parameterized as GDM may be genuine fluid components (e.g. scalar or vector fields), or \textit{effective} fluids (eg. the scalaron of $f(R)$ gravity, the Weyl fluid of DGP gravity - see \textsection\ref{section:examples}). For example, in eqn.(\ref{rescaled_general_Udelta2}) we identify the extra d.o.f. with an energy density perturbation:
\begin{equation}
\label{fluid_identify}
 \vec{\alpha}_0^{\mathbf{T}}\hat{\vec{\chi}}+\vec{\alpha}_1^{\mathbf{T}}\dot{\hat{\vec{\chi}}}=8\pi G_0 a^2\rho_E\Delta_E
 \end{equation}
 where the `hat' symbol indicates that we have folded in the necessary metric perturbations to make gauge-invariant versions of $\chi$ and $\dot\chi$. Similarly we can construct gauge-invariant $\hat{\vec{\chi}},\, \dot{\hat{\vec{\chi}}}$ and $\ddot{\hat{\vec{\chi}}}$ from the terms in eqns.(\ref{extrafields_theta})-(\ref{extrafields_Sigma}), and identify these with velocity, pressure and anisotropic stress perturbations respectively. A subscript $E$ will be used to indicate these effective perturbations.
 
 The GDM formalism then provides a way of reducing these four fluid perturbations to just two, which can then be related to the metric potentials via the perturbed conservation equations. These are \cite{Hu_GDM}:
 \begin{eqnarray}
 \dot\delta_E&=&-(1+w_E)\left(k^2\theta_E-3\dot\Phi\right)+\frac{\dot w_E\delta_E}{1+w_E}-3{\cal H}w_E\Gamma_E+\nonumber\\
 &&\hspace{-1cm}\left[\ldots\right]\Phi+\left[\ldots\right]\dot\Phi+\left[\ldots\right]\ddot\Phi+\left[\ldots\right]\Psi+\left[\ldots\right]\dot\Psi+\left[\ldots\right]\ddot\Psi\label{Hu_consv_1}
 \end{eqnarray}
 \begin{eqnarray}
 \dot\theta_E &=& -{\cal H}\left(1-3 c_{\mathrm{ad}}^2\right)\theta_E+\frac{c_{\mathrm{ad}}^2\delta_E+w_E\Gamma_E}{(1+w_E)}-\frac{2}{3}\Sigma_E+\nonumber\\
 &&\hspace{-1cm}\left[\ldots\right]\Phi+\left[\ldots\right]\dot\Phi+\left[\ldots\right]\ddot\Phi+\left[\ldots\right]\Psi+\left[\ldots\right]\dot\Psi+\left[\ldots\right]\ddot
\Psi \label{Hu_consv_2}
 \end{eqnarray} 
 where the square brackets denote combinations of the functions $B_i,\ldots K_i$. We should remember that these are really second-order equations, due to the $\dot{\hat{\vec\chi}}$ in eqn.(\ref{fluid_identify}). 
 In eqn.(\ref{Hu_consv_2}) the pressure perturbation of the effective fluid has been decomposed into an adiabatic and a non-adiabatic part:
 \begin{equation}
 \Pi_E=c_{\mathrm{ad}}^2\delta_E+w_E\Gamma_E
 \end{equation}
 We have adopted common notation by using $\Gamma_E$ to represent the dimensionless non-adiabatic pressure perturbation; this should not be confused with our metric potential $\hat\Gamma$. The adiabatic sound speed is fully determined by the equation of state parameter $w_E$:
 \begin{equation}
 \label{Hu_sound_speed}
 c_{\mathrm{ad}}^2=w_E-\frac{1}{3 {\cal H}}\frac{\dot w_E}{1+w_E}
\end{equation} 
The non-adiabatic pressure perturbation is specified by introducing a parameter $c_{\mathrm{eff}}^2$, interpreted as the sound speed of the fluid in its rest frame:
\begin{equation}
w_E\Gamma_E = \left(c_{\mathrm{eff}}^2-c^2_{\mathrm{ad}}\right)\big(\delta_E+3 {\cal H} \left( 1+w_E\big)\theta_E\right)
\end{equation}
A third and final parameter is needed to relate the anisotropic stress, $\Sigma_E$, to the velocity perturbations. This is the viscosity parameter $c_{\mathrm{vis}}^2$:
\begin{equation}
\label{Hu_viscosity_parameter}
(1+w_E)\left(\dot\Sigma_E+3{\cal H}\Sigma_E\right)-\frac{\dot w_E}{w_E}\Sigma_E=4 c_{\mathrm{vis}}^2 k^2 \theta_E
\end{equation}
By combining eqns.(\ref{Hu_sound_speed})-(\ref{Hu_viscosity_parameter}) with eqns.(\ref{Hu_consv_1}) and (\ref{Hu_consv_2}) we can obtain two equations relating any two of the fluid perturbations to the two metric potentials (although we note that the presence of the $\dot\Sigma_E$ term in eqn.(\ref{Hu_viscosity_parameter}) might make this step non-trivial). These expressions will contain the three GDM parameters $\{w_E, c_{\mathrm{eff}}^2, c_{\mathrm{vis}}^2\}$. An equivalent three-parameter framework was studied in the context of dark energy in \cite{Koivisto_Mota}, and constraints from current and future data sets were investigated in \cite{Mota_2007}.

We have already mentioned the degeneracy bewteen modifications to gravity and fluid components in the zeroth-order field equations \cite{Hu_Eisenstein_1999}. An explicit example of this is presented in \cite{Banados_Ferreira_Skordis} for the case of Eddington-Born-Infeld gravity. We stress that the effective fluids representing the extra d.o.f. at the background and perturbed levels need not have the same properties. Indeed, it is most likely that they will be different. If they are not, we have no way of deciding whether dark energy is really a modification to gravity or a dark fluid, even using observations that reflect the rate of growth of structure \cite{Kunz_Sapone}.

As described above, for theories with two or less extra d.o.f. the GDM prescription is not strictly needed; the system of equations is already closed. But if we wish to constrain modified gravity in a model-independent way we cannot make assumptions about the number of new d.o.f. introduced. The GDM approach allows us to obtain model-independent closure relations at the expense of introducing three new parameters, which would need to be constrained in a MCMC analysis.

\section{Examples}
\label{section:examples}
Thus far our discussion of the PPF parameterization has been purely formal. Studying some specific cases of gravitational theories should help to consolidate the ideas outlined in this paper. The four examples presented here are chosen to represent some common classes of theories of modified gravity; a comprehensive review of many other theories is presented in \cite{MG_report}. A simple scalar-tensor theory and Einstein-Aether theory represent theories that introduce additional fields to GR (scalar and vector fields respectively). Theories with additional tensor fields such as Eddington-Born-Infeld gravity \cite{Banados2008,Banados_Ferreira_Skordis} and bimetric theories \cite{Milgrom_bimetric_MOND} also belong to this broad category. $f(R)$ gravity is studied as an example of a higher-derivative theory; Ho\v{r}ava-Liftschitz gravity \cite{HL1, HL2, Sotiriou_HL_status_report} and Galileon theories \cite{Nicolis_Galileons, Trodden_Galileons} also fall into this class. Our final example is Dvali-Gabadazze-Porrati gravity (DGP) \cite{DGP2000,Deffayet2002}, which we study as a representative higher-dimensional theory. Whilst DGP itself is now disfavoured by observations, it incorporates features common to other braneworld theories \cite{MaartensKoyama,add1998} such as Randall-Sundrum models \cite{RandallSundrum1, RandallSundrum2} and cascading gravity \cite{deRham_CascadingGravity}. 

For each theory we will consider the extent to which the gravitational field equations can be modelled by the PPF parameterization. Is it possible to describe such a rich variety of theories using only two functions? The usual arena for PPF is the quasistatic limit, in which time derivatives of perturbations can be neglected relative to spatial derivatives. This is the dominant regime for measures of the rate of structure growth, such as weak lensing and peculiar velocity surveys. However, we are also interested in using constraints from the Integrated Sachs-Wolfe effect, which requires consideration of scales above the quasistatic regime. There has been much work recently highlighting the importance of a correct relativistic treatment of large scales, for both the theoretical and observed matter power spectrum \cite{Yoo, Bonvin_Durrer, Challinor_Lewis, Bruni_et_al_2011}. We will consider whether the PPF parameterization can be extended to this regime. 

\subsection{Scalar-tensor theory}
\label{subsection:scalar_tensor}

Scalar-tensor theories, first considered by Jordan, Brans and Dicke \cite{BransDicke}, modifiy GR by introducing a scalar field which couples to the Ricci scalar in the gravitational action. The concept is closely linked to that of quintessence, in which a scalar field is used a dark energy fluid-type component but without the explicit non-minimal coupling to the Ricci scalar. Particle physics is in no short supply of candidate scalar fields, and the reduction of higher-dimensional theories to effective four-dimensional field theories also gives rise to candidate scalars (moduli). However, finding a field with exactly the right properties to account for dark energy has proved difficult. There is an obvious aesthetic appeal in connecting the scalar field with the inflaton; however, they need not \textit{a priori} be the same field, and introducing such a connection creates further obstacles to constructing viable models.

In general, three functions are required to specify a scalar-tensor theory: the coupling to the Ricci scalar, \mbox{$F(\phi)$}, a potential for the scalar field, \mbox{$U(\phi)$}, and a function multiplying the kinetic term of the scalar, \mbox{$Z(\phi)$}. However, it is possible to reduce \mbox{$F(\phi)$} and \mbox{$Z(\phi)$} to just one function through a field redefinition, resulting in \mbox{$F(\phi)=\phi$} and \mbox{$Z(\phi)=\omega (\phi)/\phi$} \cite{Esposito_Farese}. The choice $\omega(\phi)=$ constant is termed a Brans-Dicke theory, which recovers GR plus a cosmological constant in the limit $\omega \rightarrow \infty, \,U(\phi)\rightarrow \Lambda$. Measurements from the Cassini spacecraft constrain \mbox{$\omega \gtrsim 40,000$} (2$\sigma$ limits) in the solar system \cite{Bertotti2003}.

A generic property of scalar-tensor theories is that they result in a time-dependent gravitational `constant'. This is precisely one of the features of the PPF formalism (through $\tilde g$ in our parameterization), which gives hope that scalar-tensor gravity might be fully accommodated by PPF. For simplicity we will focus on a scalar-tensor theory with variable coupling $\omega(\phi)$ and no potential,
working in the conformal Newtonian gauge. The action in the Jordan frame is:
\begin{equation}
\label{ST_action}
S = \frac{1}{16\pi}\int \mathrm{d}^4x\sqrt{-g} \left[\phi R -\frac{\omega(\phi)}{\phi} (\nabla\phi)^2  \right]+S_m[\psi^{(i)}, g_{\mu\nu}]
\end{equation}
where $S_m[\psi^{(i)}, g_{\mu\nu}]$ is the matter action and $\psi^{(i)}$ the matter fields. Varying this action with respect to the metric yields the gravitational field equations \cite{Nagata}:
\begin{eqnarray}
\label{ST_field_eqn}
G_{\mu\nu} &=& \frac{8\pi G_0}{\phi}\, T_{\mu\nu}+\frac{\omega(\phi)}{\phi^2}\left( \nabla_{\mu}\phi \nabla_{\nu}\phi-\frac{1}{2}g_{\mu\nu}(\nabla\phi)^2\right) \nonumber \\
&&+\frac{1}{\phi}\left(\nabla_{\mu}\nabla_{\nu}\phi-g_{\mu\nu}\,\square\,\phi\right)
\end{eqnarray}
where $\square = g^{\mu\nu}\nabla_{\mu}\nabla_{\nu}$. $G_{\mu\nu}$ is the usual Einstein tensor of GR and the scalar field has been rescaled so that it is dimensionless, $\phi\rightarrow\phi/G_0$. 
In a smooth, unperturbed FRW universe this gives us the background equations (c.f. eqns.(\ref{E_F}) and (\ref{E_R})):
\begin{eqnarray}
E_F&=&\frac{8\pi G_0}{\phi}a^2\sum_i\rho_i+\frac{1}{2}\omega(\phi)\frac{\dot\phi^2}{\phi^2}-3{\cal H}\frac{\dot\phi}{\phi}\label{st_background_1}\\
E_R&=&\frac{8\pi G_0}{\phi}a^2 \sum P_i+\frac{1}{2}\omega(\phi)\frac{\dot\phi^2}{\phi^2}+\frac{\ddot\phi}{\phi}+{\cal H}\frac{\dot\phi}{\phi}\label{st_background_2}
\end{eqnarray}
Variation of the action with respect to $\phi$ gives the equation of motion for the scalar field \cite{Nagata}:
\begin{equation}
\label{ST_eom}
\square\phi=\frac{1}{2\omega (\phi)+3}\left(8\pi G_0 a^2 T^{\mu}_{\mu}-\frac{\mathrm{d}\omega (\phi)}{\mathrm{d}\phi}(\nabla\phi)^2\right)
\end{equation}
Rewriting eqn.(\ref{ST_field_eqn}) in the form of eqn.(\ref{einstein}) indicates that the form of $U_{\mu\nu}$ must be:
\begin{eqnarray}
\label{ST_U_zero}
U_{\mu\nu} &=& G_{\mu\nu}(1-\phi)+\frac{\omega(\phi)}{\phi}\left( \nabla_{\mu}\phi \nabla_{\nu}\phi-\frac{1}{2}g_{\mu\nu}(\nabla\phi)^2\right) \nonumber \\
&&+\nabla_{\mu}\nabla_{\nu}\phi-g_{\mu\nu}\,\square\,\phi
\end{eqnarray}
Linearly perturbing this expression (and raising an index) will give us $U_{\Delta}, U_{\Theta}, U_P$ and $U_{\Sigma}$. We obtain:
\begin{eqnarray}
U_{\Delta} &=& E_{\Delta} (1-\phi) +\dot\Phi\, [3\dot\phi]+\Psi\left[6 {\cal H}\dot\phi-\omega(\phi)\frac{\dot\phi^2}{\phi}\right]\label{ST_UDelta}\nonumber\\
&+&\delta\phi \left[\frac{1}{2}\frac{\mathrm{d}\omega(\phi)}{\mathrm{d}\phi}\,\frac{\dot\phi^2}{\phi}-\frac{1}{2}\omega(\phi)\frac{\dot\phi^2}{\phi^2}-k^2-3 {\cal H}^2\right]\nonumber\\
&+&\dot{\delta\phi}\left[-3{\cal H}+\omega(\phi)\frac{\dot\phi}{\phi}\right]\\
U_{\Theta}&=&E_{\Theta}(1-\phi)+\delta\phi\left[\omega(\phi)\frac{\dot\phi}{\phi}-{\cal H}\right]+\dot{\delta\phi}+\Psi[-\dot\phi]\label{ST_Utheta} \\ 
U_P &=& E_P (1-\phi)+\dot\Phi\,\left[-6\dot\phi\right]\nonumber\label{ST_UP}\\
&+&\Psi \left[-6\ddot\phi-6{\cal H}\dot\phi-3\omega(\phi)\frac{\dot\phi^2}{\phi}\right]+\dot\Psi\,[-3\dot\phi]\nonumber\\
&+&\delta\phi\left[-\frac{3}{2}\omega(\phi)\frac{\dot\phi^2}{\phi^2}+\frac{3}{2} \frac{\mathrm{d}\omega(\phi)}{\mathrm{d}\phi}\,\frac{\dot\phi^2}{\phi}+3{\cal H}^2+6\dot{\cal H}+2k^2\right]\nonumber\\
&+&\dot{\delta\phi}\left[3{\cal H}+3\omega(\phi)\frac{\dot\phi}{\phi}\right]+3\,\ddot{\delta\phi}\\
U_\Sigma &=&E_{\Sigma}(1-\phi)+\delta\phi\label{ST_USigma} 
\end{eqnarray}
Specialising to the conformal Newtonian gauge (in which $V$=0) means that we lose the time derivatives in eqns.(\ref{phi_def}) and (\ref{psi_def}). Hence the appearance of $\Psi$ in $U_{\Delta}$ and $U_{\Theta}$ above gives us no cause for concern; the Bianchi identities will remain second-order equations. If we had kept to a general gauge additional terms in $V, \epsilon$ and $\nu$ would be present in the above expressions, ensuring that $\hat\Psi$ only appeared within the combination $\dot{\hat\Phi}+{\cal H}\hat\Psi$ and that any $\ddot a/a$ terms cancel. The (lengthy) corresponding expressions for a general gauge are displayed in \makebox{Appendix \ref{app:gauge_invariance}}. There we also demonstrate that the general-gauge expressions still obey the constraints of yielding second-order, gauge-invariant perturbed field equations.

We  can see that eqn.(\ref{ST_UDelta}) has the form indicated in eqn.(\ref{rescaled_general_Udelta2}); 
in this particular case we can pick out the coefficient functions:
\begin{eqnarray}
 A_0&=&-2(1-\phi)\nonumber\\
 A_1&=&\frac{3\dot\phi}{k}-6{\cal H}_k(1-\phi)\nonumber \\
 F_0&=&-6{\cal H}_k^2(1-\phi)+6\frac{{\cal H}_k}{k}\dot\phi-\frac{\omega(\phi)}{k^2}\frac{\dot\phi^2}{\phi}\nonumber\\
 F_1&=&0\nonumber\\
 \alpha_0&=&\frac{1}{2k^2}\frac{\mathrm{d}\omega(\phi)}{\mathrm{d}\phi}\,\frac{\dot\phi^2}{\phi}-\frac{1}{2k^2}\omega(\phi)\frac{\dot\phi^2}{\phi^2}-1-3 {\cal H}_k^2\nonumber\\
 \alpha_1&=&-3{\cal H}_k+\frac{\omega(\phi)}{k}\frac{\dot\phi}{\phi}
 \end{eqnarray}
 Similarly one can read off the coefficients $\beta_i,\ldots\varepsilon_i$ and $B_i,\ldots K_i$ for scalar-tensor theory by matching eqns.(\ref{extrafields_theta})-(\ref{extrafields_Sigma}) with eqns.(\ref{ST_Utheta})-(\ref{ST_USigma}).
 
 Using the expressions for $U_{\Delta}$ and $U_{\Theta}$ in eqns.(\ref{einstein_perturbed}) we can form the modified (Fourier-space) Poisson equation. Some of the terms in $\delta\phi$ and $\delta\dot\phi$ can be combined to form the gauge-invariant density perturbation of a scalar field:
\begin{equation}
\rho_{\phi}\Delta_{\phi}=\delta\rho_{\phi}+3{\cal H}\left(\rho_{\phi}+P_{\phi}\right)\theta_{\phi}=\dot\phi\left(\delta\dot\phi-\dot\phi\Psi+3{\cal H}\,\delta\phi\right)
\label{scalar_field_gi_perturbation} 
\end{equation}
The Poisson equation for scalar-tensor gravity is then:
\begin{eqnarray}
\label{ST_Poisson}
-2k^2\Phi&=&\frac{8\pi G_0 a^2}{\phi}\left(\sum_i\rho_i\Delta_i+\rho_{\phi}\frac{\omega(\phi)}{\phi}\Delta_{\phi}\right)\\
&&+3\frac{\dot\phi}{\phi}\left[\dot\Phi+{\cal H}\Psi\right]\nonumber\\
&&-\frac{\delta\phi}{\phi}\left[k^2+6{\cal H}^2+\frac{1}{2}\frac{\dot\phi^2}{\phi}\left(\frac{\omega(\phi)}{\phi}-\frac{\mathrm{d}\omega(\phi)}{\mathrm{d}\phi}\right)\right] \nonumber 
\end{eqnarray}
 $\Delta_i$ is a gauge-invariant density perturbation to matter (CDM, baryons, radiation), and we have redefined the scalar field so as to pull a factor of $8\pi G_0 a^2$ out of the second term. We have chosen to write the Poisson equation in this form because it delineates the extra terms that arise in a scalar-tensor theory compared to uncoupled quintessence. We reach a quintessence-like limit by setting $\omega(\phi)=\phi$ and removing the $\phi\,$-$R$ coupling in the action. As a result, the last two lines of eqn.(\ref{ST_Poisson}) do not arise and the prefactor is just $8\pi G_0 a^2$ (because we don't rescale $\phi\rightarrow\phi/G_0$). It is the last two lines that offer the distinction between quintessence and scalar-tensor gravity.
 The slip relation is:
\begin{equation}
\Phi-\Psi=\frac{\delta\phi}{\phi}\label{ST_slip}
\end{equation}

It is useful to consider some simplifying limits of the above expressions. The `smooth' limit, in which  the perturbations of the scalar field are negligible relative to the matter perturbations, gives us:
\begin{equation}
\label{ST_Poisson_smooth}
-2k^2\Phi=8\pi \frac{G_0}{\phi} a^2 \sum_i\rho_i\Delta_i+3\frac{\dot\phi}{\phi}\left[\dot\Phi+{\cal H}\Psi\right]
\end{equation}
The second term on the right-hand side cannot be absorbed into the first term without giving $G_{\mathrm{eff}}$ an undesired environmental dependence. 

However, if we take the quasistatic limit of eqn.(\ref{ST_Poisson}) -- that is, we neglect time derivatives of perturbation variables and take $k \gg {\cal H}$ -- we find that the modifications to the Poisson equation can indeed be repackaged as a modified gravitational constant. In this limit it is possible to write the relation between the two metric potentials as $\Phi=\eta_{\mathrm{slip}}\Psi$; e.g. for the choice $\omega(\phi)=\phi$ the form of $\eta_{\mathrm{slip}}$ is particularly simple \cite{Amendola_WL, Schimd}:
\begin{equation}
\label{ST_QS_slip}
\eta_{\mathrm{slip}}= \frac{\phi+1}{\phi}\\
\end{equation}
Combining eqns.(\ref{ST_Poisson}) and (\ref{ST_slip}) with the slip relation, we obtain:
\begin{eqnarray}
&&-k^2\Phi = 8\pi \frac{G_0}{\phi} a^2 \left(\sum_i\rho_i\Delta_i+\frac{\omega(\phi)}{\phi}\rho_{\phi}\Delta_{\phi}\right)\times\label{ST_QS_Poisson}\\
&&\hspace{-6mm}\Bigg[1+\frac{1}{\eta_{\mathrm{slip}}}-\left(1-\frac{1}{\eta_{\mathrm{slip}}}\right)\frac{\dot\phi^2}{2k^2\phi}\left[\frac{\omega(\phi)}{\phi}-\frac{\mathrm{d}\omega(\phi)}{\mathrm{d}\phi}\right]\nonumber\\
&&\hspace{5.9cm}+\,\frac{3\cal H}{k^2\eta_{\mathrm{slip}}}\frac{\dot\phi}{\phi}\Bigg]^{-1}\nonumber
\end{eqnarray}
where the expression in square brackets gives the time- and scale-dependence of Newton's gravitational constant.

As mentioned earlier, it is important that we also consider the (super)horizon-scale limit of these theories for correct treatment of their predicted effects on the matter power spectrum and large-angle CMB power spectrum. Taking the limit \mbox{$k\ll{\cal H}$} allows us to neglect the $k^2 \delta\phi/\phi$ term of eqn.(\ref{ST_Poisson}), but there are no other obvious simplifications unless $\delta\phi$ is negligible on these scales. It seems that in order to cope with the superhorizon limit we need a parameterization that allows for additive modifications to the Poisson equation, as well as a modified $G_{\mathrm{eff}}$. Of course, there is no barrier to using the standard PPF format to model \textit{part} of the modifications,  but we need to remember that the correspondence between parameterization and theory would no longer be exact on large scales.

\subsection{$f(R)$ gravity}
\label{subsection:f_r}
Another commonly-studied theory is $f(R)$ gravity, for which the action is a general function of the Ricci scalar, $f(R)$; see \cite{Sotiriou_fR} for a detailed review. Two approaches to $f(R)$ gravity are possible. In the metric formulation the affine connection
$\Gamma^{\rho}_{\mu\nu}$ is defined in terms of the metric components in the usual way, and gravitational field equations are obtained by varying the metric with respect to $g_{\mu\nu}$ only. In the Palatini formulation of $f(R)$ gravity the connection and the metric are treated as independent variables and the action is varied with respect them individually. One finds that in the metric formulation there is a propagating scalar degree of freedom \mbox{$f_R=\partial f(R)/\partial R$}. Since the Ricci scalar is constructed from second derivatives of the metric, the kinetic term of the scalar degree of freedom contains fourth-order derivatives and hence metric $f(R)$ gravity corresponds to a higher-order theory. 

The presence of a scalar degree of freedom within $f(R)$ theories can be made explicit by applying a conformal transformation that maps $f(R)$ gravity onto a scalar-tensor theory. Metric and Palatini $f(R)$ gravity map onto Brans-Dicke theories with $\omega=0, \,-\frac{3}{2}$ respectively \cite{Sotiriou_fR}. The scalar field arising under the conformal transformation is sometimes referred to as the `scalaron'.  In the conformally-transformed frame (the Einstein frame) the scalaron acquires a coupling to the matter fields $\psi^{(i)}$, leading to non-standard conservation equations. Hence the non-transformed frame (the Jordan frame) is regarded as the physical frame in which observations are made. 

The action in the Jordan frame is;
\begin{equation}
\label{fr_action}
S = \frac{1}{16\pi G_0}\int \mathrm{d}^4x \sqrt{-g}\, f(R) +S_m[\psi^{(i)}, g_{\mu\nu}]
\end{equation}
Variation with respect to the metric leads to the field equations:
\begin{equation} 
f_R R_{\mu\nu} -\frac{1}{2} f(R) g_{\mu\nu} - \nabla_u\nabla_v f_R +g_{\mu\nu}\square f_R = 8\pi a^2 G_0 T_{\mu\nu}
\end{equation}
This can be rewritten in the form of eqn.(\ref{einstein}) with the following expression for $U_{\mu\nu}$:
\begin{equation}
U_{\mu\nu}=R_{\mu\nu} (1-f_R)-\frac{1}{2}g_{\mu\nu}(R-f(R))+\nabla_u\nabla_v f_R -g_{\mu\nu}\square f_R 
\end{equation}
We will let $\chi$ denote the perturbation to the extra degree of freedom, ie. \mbox{$\chi=\delta(f_R)=f_{RR}\delta R$}. Perturbing the above expression \begin{eqnarray}
U_{\Delta} &=& E_{\Delta} (1-f_R) +\dot\Phi\, [3\dot f_R]+\Psi\left[6 {\cal H}\dot f_R\right]\nonumber \\
&&+\chi \left[3 \dot{\cal H}-k^2\right]+\dot\chi\left[-3{\cal H}\right] \label{fR_UDelta}\\
U_{\Theta}&=&E_{\Theta}(1-f_R)+\chi\left[-{\cal H}\right]+\dot\chi+\Psi[-\dot f_R]\label{fR_Utheta}\\ 
U_P &=& E_P (1-f_R)+\dot\Phi\,\left[-6\dot f_R\right]+\Psi \left[-6\ddot f_R-6{\cal H}\dot f_R\right]\label{fR_UP}\\
&&+\dot\Psi\,[-3\dot f_R]+\chi\left[-3\dot{\cal H}-6{\cal H}^2+2k^2\right]+\dot\chi\left[3{\cal H}\right]+3\,\ddot\chi\nonumber\\
U_\Sigma &=&E_{\Sigma}(1-f_R)+\chi\label{fR_USigma}
\end{eqnarray}
The similarity between the sets of eqns.(\ref{ST_UDelta})-(\ref{ST_USigma}) and (\ref{fR_UDelta})-(\ref{fR_USigma}) is immediately apparent, suggesting the identification of $f_R$ as the scalaron. However, in $f(R)$ gravity the `extra' d.o.f, $\chi$, can be expressed in terms of metric potentials.   $\dot\chi$ and $\ddot\chi$ are given by the expressions:
\begin{eqnarray}
\dot\chi &=& f_{3R} \dot R \,{\delta R}+f_{RR}\dot{(\delta R)}\\
\ddot\chi&=&\delta R\, (f_{4R}\dot R^2+f_{3R}\ddot R)+2 f_{3R}\dot R \dot{(\delta R)}+f_{RR}\ddot{(\delta R)}
\end{eqnarray}
where $f_{4R}$ indicates the fourth derivative of $f$ with respect to $R$ etc., and 
\begin{widetext}
\begin{eqnarray}
a^2\delta R&=& -4 k^2\Phi-18{\cal H}\dot\Phi-6 \ddot\Phi-2\left(6 \dot{\cal H} + 6 {\cal H}^2 - k^2\right)\Psi-6{\cal H}\dot\Psi\label{f_R_deltaR}\\
a^2(\dot{\delta R})&=&-2a^2{\cal H} \delta R-\dot\Phi \left(4 k^2 +18\dot{\cal H}\right) - 18{\cal H}\ddot\Phi - 6\Phi^{(3)} -12\Psi\left(\ddot{\cal H}+ 2{\cal H}\dot{\cal H}\right) - \dot\Psi \left(18\dot{\cal H} + 12 {\cal H}^2 - 2 k^2\right)- 6 {\cal H} \ddot\Psi\\
a^2(\ddot{\delta R})&=&-\delta R\, a^2 \left(2\dot{\cal H}+4{\cal H}^2\right)-4a^2{\cal H}(\dot{\delta R})-18\ddot{\cal H} \dot\Phi - \ddot\Phi \left(36 \dot{\cal H} +4k^2\right) - 18 {\cal H} \Phi^{(3)}- 6 \Phi^{(4)} - 12 \Psi \left({\cal H}^{(3)} + 2 \dot{\cal H}^2 + 2 {\cal H}\ddot{\cal H}\right) \nonumber\\
&&- \dot\Psi \left(30 \ddot{\cal H} + 48 \dot{\cal H} {\cal H}\right)- \ddot\Psi \left(24 \dot{\cal H} + 12 {\cal H}^2-2k^2\right) - 6 {\cal H}\Psi^{(3)}
  \end{eqnarray}
  \end{widetext}
The coefficients corresponding to the general expressions in \textsection\ref{section:extra_fields} can be computed. For example, matching onto eqn.(\ref{rescaled_general_Udelta2}):
\begin{eqnarray}
A_0&=&-2\left(1-f_R\right)-4\frac{f_{RR}}{a^2}\left(3\dot{\cal H}-k^2+6{\cal H}^2\right)\nonumber\\
&&\hspace{5.5cm}+12{\cal H}\frac{f_{3R}}{a^2}\dot R\nonumber\\
A_1&=&-6{\cal H}_k\left(1-f_R\right)+3\frac{\dot f_R}{k}+6\frac{f_{RR}{\cal H}_k}{a^2}\left(5k^2-18{\cal H}^2\right)\nonumber\\
&&+54\frac{f_{3R}}{ka^2}{\cal H}^2\dot R \nonumber\\
A_2&=&6\frac{f_{RR}}{a^2}\left(k^2+3{\cal H}^2-3\dot{\cal H}\right)+18\frac{f_{3R}}{a^2}{\cal H}\dot R \nonumber\\
A_3&=&18k\frac{f_{RR}}{a^2}{\cal H} \nonumber
\end{eqnarray}
\begin{eqnarray}
F_0&=&-6{\cal H}_k^2\left(1-f_R\right)+6\frac{{\cal H}_k}{k}\dot f_R+6\frac{f_{RR}}{a^2}\Bigg[-12{\cal H}^2{\cal H}_k^2\nonumber\\
&&-6{\cal H}_k^2\dot{\cal H}-6\dot{\cal H}_k^2+6{\cal H}_k\ddot{\cal H}_k+4{\cal H}^2+3\dot{\cal H}-\frac{k^2}{3}  \Bigg]\nonumber\\
&&+6\frac{f_{3R}}{a^2}{\cal H}\dot R\left[6{\cal H}_k^2+6\frac{\dot{\cal H}_k}{k}-1\right] \nonumber\\
F_1&=&18\frac{{\cal H}_k}{a^2}\left(2f_{RR}\dot{\cal H}+f_{3R}{\cal H}\dot R\right) \nonumber\\
F_2&=&18{\cal H}^2\frac{f_{RR}}{a^2} \nonumber\\
\alpha_0^{(i)}&=&\alpha_1^{(i)}=\alpha_2^{(i)}=\alpha_3^{(i)}=0
\end{eqnarray}
The slip relation and modified Poisson equation for $f(R)$ gravity become:
\begin{eqnarray}
\Phi-\Psi&=&\frac{\chi}{f_R}\label{f_R_slip}\\
-2k^2\Phi&=&8\pi \frac{G_0}{f_R}a^2 \sum_i \rho_i\Delta_{i}+3\frac{\dot f_R}{f_R}\left[\dot\Phi+{\cal H}\Psi\right]\nonumber\\
&&-\frac{\chi}{f_R}\left[\frac{3}{2}E+k^2\right]\label{f_R_Poisson}\nonumber\\
\end{eqnarray}
Again we see that these expressions are largely similar to the scalar-tensor case with $\omega(\phi)=0$, $f_R$ replacing $\phi$ and $\chi \equiv \delta\phi$. However, we note that the similarities are aesthetic only; the equivalence between scalar-tensor theory and $f(R)$ gravity is only formally realised under a conformal transformation as described above. $\delta\phi$ represents a perturbation to a new field that is genuinely additional to GR, whereas $\chi$ is really a placeholder for the combination of metric perturbations in eqn.(\ref{f_R_deltaR}). 

In the `smooth' limit $\chi\rightarrow 0$ we obtain the analogy of eqn.(\ref{ST_Poisson_smooth}):
\begin{equation}
\label{fR_Poisson_smooth}
-2k^2\Phi=8\pi \frac{G_0}{f_R} a^2 \sum_i\rho_i\Delta_i+3\frac{\dot f_R}{f_R}\left[\dot\Phi+{\cal H}\Psi\right]
\end{equation}
with the same second background-dependent term as the scalar-tensor case. 

For measures of late-time structure  growth we are predominantly interested in the quasistatic regime. Using $\chi=f_{RR} \delta R$ and eqn.(\ref{f_R_deltaR}) we can replace $\chi$ in the Poisson equation and slip relation by sequence of the metric potentials and their derivatives, just as we laid out in eqns.(\ref{Udelta})-(\ref{Usigma}). For example, in the quasistatic limit the slip relation becomes:
\begin{equation}
\label{f_R_QS_slip_me}
\Phi-\Psi = \frac{f_{RR}}{f_R a^2}\left(2(-6\dot{\cal H}+k^2)\Psi-4k^2\Phi\right)
\end{equation}
This matches the result of Pogosian \& Silvestri \cite{Pogosian_Silvestri} if we further neglect the $\dot{\cal H}$ term. They expressed the slip parameter of $f(R)$ gravity in the quasistatic limit as:
\begin{equation}
\label{f_R_QS_slip_PS}
\eta_{\mathrm{slip}}=\frac{3+2Q}{3+4Q}
\end{equation}
where $Q$ is defined as the squared ratio of the Compton wavelength of the scalaron to the physical wavelength of a mode:
\begin{equation}
\label{scalaron_Compton}
Q=3\frac{k^2}{a^2}\frac{f_{RR}}{f_R} \approx \left(\frac{\lambda_C}{\lambda}\right)^2
\end{equation}
In this limit the Poisson equation can be rewritten as:
\begin{eqnarray}
\label{fR_Poisson_QS}
-k^2\Phi&=& 8\pi \frac{G_0}{f_R}a^2\sum_i\rho_i\Delta_i  \times \\
&&\hspace{-0.5cm}\left[1+\frac{1}{\eta_{\mathrm{slip}}}\left(1+3\frac{\dot{f_R}}{f_R} \frac{\cal H}{k^2}\right)-\frac{3}{2}\frac{E}{k^2}\left(1-\frac{1}{\eta_{\mathrm{slip}}}\right)\right]^{-1} \nonumber
\end{eqnarray}
The term in square brackets controls the time- and scale-dependence of the effective Newton's constant.

It is no surprise to find that $f(R)$ gravity behaves similar to scalar-tensor theory in the superhorizon limit. We can neglect $k^2$ in eqn.(\ref{f_R_Poisson}) and eliminate $\chi$ using eqn.(\ref{f_R_slip}), but the resulting expression will still have terms in $\Psi$  and $\dot\Phi$ on the right-hand side that can not be written as a $G_{\mathrm{eff}}$. The conclusion is analogous to that of \textsection\ref{subsection:scalar_tensor}: the mapping between a ($G_{\mathrm{eff}},\eta_{\mathrm{slip}}$) parameterization is exact in the quasistatic limit, but only approximate in the large-scale regime.

\subsection{Einstein-Aether theory}
\label{subsection:EA}
The emergence of string theory as a candidate theory of quantum gravity leads to the possibility that spacetime coordinates are non-commutative \cite{Carroll_Harvey_2001}. Under these circumstances Lorentz symmetry may be violated. Hence much effort has been invested in exploring ways in which Lorentz violation can be implemented without marring the key successful features of GR, such as general covariance. A minimal way to do this is to introduce a vector field into the action, which defines a preferred reference frame at every point in spacetime. In Minkowski spacetime one can simply introduce a constant vector field into the background, but in a curved spacetime this is not possible; the Lorentz-violating vector field must be promoted to a dynamical field derived from a generally covariant action.
If invariance under three-dimensional spatial rotations is preserved then the vector field must be time-like. Einstein-Aether theories  \cite{JacobsonMattingly} introduce such a vector field (`the aether') of unit length, which is constrained not to vanish so that Lorentz violation is maintained even in a vacuum. The unit, time-like nature of the vector field is enforced through means of a Lagrange multiplier in the action.  

TeVeS  \cite{Bekenstein_TeVeS_2004} is a well-known example of another theory that contains Lorentz-violating vector fields. When formulated in the Einstein frame, TeVeS introduces both a unit, timelike vector field and a scalar field, and employs a free function to ensure that it reduces to Modified Newtonian Dynamics (MOND) \cite{Milgrom1983} in the non-relativisitc limit. Upon transformation to the Jordan frame 
- in which particles follow geodesics of the metric - the unit length of the vector field is not preserved. It was shown by Zlosnik \cite{Zlosnik_TeVeS} that the scalar field present in the Einstein frame is absorbed by the vector field in the Jordan frame, and dynamically determinines the modulus of the vector. Hence TeVeS is equivalent to an Einstein-Aether theory in which the aether field has variable length. Our study of the compatibility of Einstein-Aether theory with the PPF framework therefore has implications for TeVeS also.

The most general action for Einstein-Aether theories is:
\begin{equation}
\label{EA_action}
S = \frac{1}{2\kappa^2}\int \mathrm{d}^4x \sqrt{-g}\, \left[R+{\cal L}_{EA}(g^{\mu\nu}, A^{\mu})\right] +S_m[\psi^{(i)}, g_{\mu\nu}]
\end{equation}
where the aether Lagrangian is 
\begin{equation}
{\cal L}_{EA}\left(g^{\mu\nu}, A^{\mu}\right)=M^2 {\cal F(K)}+\lambda(A^{\alpha}A_{\alpha}+1)
\end{equation}
$M$ has the dimensions of mass. ${\cal K}$ is a scalar formed from the kinetic term of the vector field:
\begin{eqnarray}
{\cal K}&=&M^{-2} {\cal K} ^{\alpha\beta}_{\gamma\sigma}\nabla_{\alpha}A^{\gamma}\nabla_{\beta}A^{\sigma}\nonumber\\
{\cal K} ^{\alpha\beta}_{\gamma\sigma}&=&c_1 g^{\alpha\beta}g_{\gamma\sigma}+c_2\delta^{\alpha}_{\gamma}\delta^{\beta}_{\sigma}+c_3\delta^{\alpha}_{\sigma}\delta^{\beta}_{\gamma} \label{EA_kappa_tensor}
\end{eqnarray}
where $c_i$ are dimensionless constants. Barbero \& Villase$\tilde{\mathrm{n}}$or \cite{Barbero_Villasenor} have identified special choices of $c_i$ for which Einstein-Aether theory becomes equivalent to GR under a field redefinition. More generally, the $c_i$ must obey certain restrictions if the linearized field equations are to be hyperbolic -- hence admitting a well-posed initial value problem -- and exclude superluminal propagation of aether perturbations and gravitational waves \cite{Zlosnik_structure_growth}. In general there is potentially a fourth term in eqn.(\ref{EA_kappa_tensor}), but in the case of purely spin-0 perturbations this can be absorbed by suitable redefinitions of $c_1$ and $c_3$. We will assume there is no direct coupling between matter and the aether; they interact only gravitationally. 

The gravitational field equations obtained by varying the Einstein-Aether action with respect to the metric can be written in the form of eqn.(\ref{einstein}), where the modification tensor $U_{\mu\nu}$ is effectively the stress-energy tensor of the aether:
\begin{eqnarray}
\label{EA_Ep_tensor}
U_{\mu\nu}&=&\nabla_{\sigma}\left({\cal F}_K\left(J_{(\mu}^{\;\;\;\sigma}A_{\nu)})-J^{\sigma}_{\;\;(\mu}A_{\nu)}-J_{(\mu\nu)}A^{\sigma}\right)\right)\nonumber\\
&&-{\cal F}_K Y_{\mu\nu}+\frac{1}{2}g_{\mu\nu}{\cal F}+\lambda A_{\mu}A_{\nu}
\end{eqnarray}
where round brackets around subscripts denotes symmetrization with weight 1/2. We use the notation ${\cal F}_K=d{\cal F}/d{\cal K}$, and the following definitions:
\begin{eqnarray}
J^{\sigma}_{\;\;\mu}&=& {\cal K} ^{\sigma\beta}_{\mu\gamma}\nabla_{\beta}A^{\gamma} \label{EA_J_def}\\
Y_{\mu\nu}&=&c_1 \left(\nabla^{\alpha}A_{\mu}\nabla_{\alpha}A_{\nu}-\nabla_{\mu}A^{\alpha}\nabla_{\nu}A_{\alpha} \right)\label{EA_Y_def}
\end{eqnarray}
Varying the action with respect to the aether field gives the equation of motion:
\begin{equation}
\label{EA_background_vecfield_eqn}
\nabla_{\mu}\left({\cal F}_KJ^{\mu}_{\;\;\;\nu}\right)=2\lambda A_{\nu}
\end{equation}
Finally, varying the action with respect to the Lagrange multiplier $\lambda$ gives the constraint $A^{\mu}A_{\mu}=-1$. The requirement of a spatially isotropic background fixes $A^{\mu}=(1,0,0,0)$. Using this in eqn.(\ref{EA_background_vecfield_eqn}) enables us to solve for the Lagrange multiplier:
\begin{equation}
\label{EA_background_Lagrange_multiplier}
\lambda=-\frac{1}{2}A^{\mu}\nabla_{\nu}\left({\cal F}_K J^{\nu}_{\;\;\mu}\right)
\end{equation}
which can be substituted back into eqn.(\ref{EA_Ep_tensor}) to eliminate $\lambda$. Defining $\alpha=c_1+3c_2+c_3$, the resulting zeroth-order field equations are:
\begin{eqnarray}
[1-{\cal F}_K\alpha]{\cal H}^2+\frac{1}{6}{\cal F} M^2a^2&=&\frac{8\pi G_0 a^2}{3}\sum_i\rho_i\label{EA_background_1}\\
&&\hspace{-4.6cm}-\left[1-2\alpha{\cal F}_K\right]{\cal H}^2-2\dot{\cal H}\left[1-\frac{1}{2}\alpha{\cal F}_K\right]\nonumber\\
+[\dot{\cal F}_K-\frac{1}{2}]{\cal F}M^2&=&8\pi G_0 a^2 \sum_i P_i\label{EA_background_2}\\
\nabla_{\mu}A^{\mu}&=&3{\cal H}\label{EA_background_3}         
\end{eqnarray}
and the kinetic scalar $\cal K$ simplifies to
\begin{equation}
 \label{EA_background_K}
 {\cal K}=3\alpha\frac{{\cal H}^2}{M^2}
 \end{equation}
The first two equations above can be written in the form of eqns.(\ref{E_F}) and (\ref{E_R}) with the identifications
\begin{eqnarray}
a^2 X &=&3{\cal F}_K\alpha{\cal H}^2-\frac{1}{2}a^2{\cal F}M^2\label{EA_X}\\
a^2 Y &=&-2{\cal F}_K\alpha{\cal H}^2+\frac{1}{2}a^2{\cal F}M^2-\alpha\left({\cal F}_K {\cal H}\right)^{\dot{}}\label{EA_Y}
\end{eqnarray}  

For simplicity we will treat only the \textit{linear} Einstein-Aether theory, in which $\cal F(K)=K$. In this particular instance eqns.(\ref{EA_background_1}) and (\ref{EA_background_2}) can alternatively be rewritten (making use of eqn.(\ref{EA_background_K}) such that they differ from the equivalent expressions in GR only through a modification to Newton's gravitational constant: 
\begin{equation}
G_0\rightarrow\frac{G_0}{1-\frac{\alpha}{2}}
\end{equation}
The situation in Einstein-Aether theory is similar to the toy model we considered in \textsection\ref{subsection:no_extra_fields_mod_background}, with the instance $\tilde g=\alpha/2$. We therefore expect the perturbations of $U_{\mu\nu}$ to look similar to eqns.(\ref{U_Delta_toy_example})-(\ref{Usigma_toy_example}). However, we have now introduced an extra field into the theory that was not present in the toy model, and this will give rise to new terms. Recall that when we applied a gauge transformation to eqn.(\ref{U_Delta_toy_example}) any terms produced by gauge-variant quantities cancelled due to the background equation (\ref{toy_example_background}). From this we can deduce that any new terms introduced by the vector field must be explicitly gauge-invariant.

Now we wish to consider linearly perturbed Einstein-Aether theory; we write the perturbations to the vector field as
\begin{equation}
\label{EA_perturbed_vector}
A^{\mu}=\left(1+Z,\;\frac{1}{a}\partial^i Q\right)
\end{equation}
Taking the linear perturbation of the constraint $A^{\mu}A_{\mu}=-1$ we find that the perturbation to the 0th component of the vector field is tied to the metric perturbations by $Z=\Xi=-\Psi$, the last equality being true in the conformal Newtonian gauge only. The perturbed equation of motion for the vector field is:
\begin{eqnarray}
\label{EA_perturbed_vecfield}
&&c_1\left(\ddot Q+k^2 Q+2{\cal H}\dot Q +2{\cal H}^2 Q+\dot\Psi+\dot\Phi+2{\cal H}\Psi\right)\nonumber\\
&&+(3 c_2+c_3)\left(k^2 Q+2{\cal H}^2 Q-\frac{\ddot a}{a}+\dot\Phi+{\cal H}\Psi\right)=0
\end{eqnarray}
 One can then find the linear perturbations of $U_{\mu\nu}$. This is a lengthy but straightforward exercise \cite{Zlosnik_structure_growth, Halle}. Using eqn.(\ref{EA_perturbed_vecfield}) to remove second-order derivatives from $U_{\Theta}$, the results are:
\begin{eqnarray}
U_{\Delta}&=&(c_1-\alpha)k^2{\cal H}Q+c_1 k^2 \left(\dot Q+\Psi\right)-3\alpha{\cal H}\left[\dot\Phi+{\cal H}\Psi\right]\label{EA_UDelta}\nonumber\\
\\
U_{\Theta}&=&\alpha\left[\dot\Phi+{\cal H}\Psi\right]+(c_1+c_2+c_3) k^2 Q\label{EA_Utheta}\\
U_P&=&\alpha k^2\left(\dot Q+2{\cal H}Q\right)\nonumber\\
&&+3\alpha\left(\ddot\Phi+2{\cal H}\dot\Phi+(2\dot{\cal H}+{\cal H}^2)\Psi+{\cal H}\dot\Psi  \right)\label{EA_UP}\\
U_{\Sigma}&=&-(c_1+c_3)\left(\dot Q+2{\cal H}Q\right)\label{EA_USigma}		
\end{eqnarray}
It is straightforward to read off the coefficients introduced in eqn.(\ref{rescaled_general_Udelta2}):
\begin{eqnarray}
A_1&=&-3\alpha{\cal H}_k \quad\quad\quad F_0=c_1-3\alpha {\cal H}^2_k\nonumber\\
\alpha_0&=&{\cal H}(c_1-\alpha)\quad\quad \alpha_1=c_1 k\nonumber\\
A_0&=&F_1=0
\end{eqnarray}
and similarly for the remaining coefficients  $\beta_i,\ldots\varepsilon_i$ and $B_i,\ldots K_i$ of eqns.(\ref{extrafields_theta})-(\ref{extrafields_Sigma}).
Note from eqn.(\ref{EA_perturbed_vecfield}) that the d.o.f. $\chi \,(\equiv Q)$ is not dimensionless, so neither are $\alpha_0$ and $\alpha_1$. 

The Poisson equation in Einstein-Aether theory is:
\begin{eqnarray}
-2k^2\Phi&=&8\pi G_0a^2\sum_i\rho_i\Delta_i\nonumber\\
&&+k^2\left({\cal H}Q(3c_1+2c_3)+c_1(\dot Q+\Psi)\right)
\label{EA_Poisson}
\end{eqnarray}
Consideration of some special cases of the $c_i$ should help us gain some understanding of this expression. Firstly we note that the slip between the metric potentials is sourced by spatial perturbations to the vector field, so in the smooth limit we recover $\Phi=\Psi$. In this limit the only modification to the Poisson equation is through a constant rescaling of Newton's constant:
\begin{equation}
\label{EA_Poisson_Qvanish}
-2k^2\Phi=8\pi \frac{G_0}{1+\frac{c_1}{2}}a^2\sum_i\rho_i\Delta_i
\end{equation}
With $Q=0$ all explicit traces of extra fields disappear from the components of $U_{\mu\nu}$ (though we ought to remember that the perturbation to the time 
component of the vector field still remains, `disguised' as the metric perturbation $\Psi$).
We note immediately that the constants renormalizing Newton's gravitational constant in the background and linear-order equations are not generally the same, being $\alpha$ and $-c_1$ respectively. If we make the choice $\alpha=-c_1$, this special case of Einstein-Aether theory can be compared to the toy model considered in \textsection\ref{subsection:no_extra_fields_mod_background}. In that example the background gravitational field equations were modified but no extra fields were present, and we assumed that the same quantity renormalized $G$ at both unperturbed and linearly perturbed order. Indeed we find that with the choices $\dot{\tilde g}=B_0=0$, $\alpha=-c_1$ and $V=\dot V=0$ (to recover the conformal Newtonian gauge) the eqns.(\ref{U_Delta_toy_example})-(\ref{U_theta_mod_background}) reproduce eqns.(\ref{EA_UDelta})-(\ref{EA_USigma}) in the limit $Q=0$.

Another case of interest is the choice $c_1+c_3=0, c_2=0$. This causes the first part of the stress-energy tensor of the aether (line 1 in eqn.(\ref{EA_Ep_tensor})) to adopt a form akin to the Maxwell tensor of electromagnetism \cite{Eling2004}. 
Under these conditions the gravitational slip again vanishes. Note that this differs from scalar-tensor and $f(R)$ gravity, where the slip could only be zero if the extra fields (treating the scalaron as an extra field) were unperturbed. 
The Poisson equation now becomes:
\begin{equation}
\label{EA_Poisson_Maxwell}
-2k^2\Phi=8\pi\frac{G_0}{1+\frac{c_1}{2}}a^2\sum_i\rho_i\Delta_i-\frac{k^2 c_1}{1+\frac{c_1}{2}}\left(\dot Q+{\cal H}Q\right)
\end{equation}
In the quasistatic limit the $\dot Q$ term can be neglected. However, the `electromagnetic condition'  has prevented us from obtaining a relation between $Q$, $\Phi$ and $\Psi$. So in this special case we are unable to package the modifications to the Poisson equation as a modified Newton's constant, even in the quasistatic regime. This is an interesting result, because thus far we have always found this to be possible. Indeed, for choices other than $c_1=-c_3$ we can write the slip relation in the usual format $\Phi=\eta_{\mathrm{slip}}\Psi$, with slip parameter:
\begin{equation}
\eta_{\mathrm{slip}}=1-4\frac{{\cal H}^2}{k^2} \frac{c_1+c_3}{c_1+c_2+c_3} \left(1-\frac{\alpha}{2}\right)
\label{EA_slip_QS}
\end{equation}
The Poisson equation is then:
\begin{eqnarray}
\label{EA_Poisson_QS}
-2k^2\Phi&=&8\pi G_0 a^2\sum_i \rho_i\Delta_i \times \\
&&\left[1+\frac{c_1}{2\eta_{\mathrm{slip}}}-\frac{3c_1+2c_3}{4(c_1+c_3)}\left(1-\frac{1}{\eta_{\mathrm{slip}}}\right)\right]^{-1}\nonumber
\end{eqnarray}
As before, the term in square brackets acts like an evolving $G_{\mathrm{eff}}$.

Whilst Einstein-Aether gravity has a similar quasistatic form to scalar-tensor gravity and $f(R)$ gravity, its large-scale behaviour is distinctly different. In the limit $k\rightarrow 0$ eqn.(\ref{EA_Poisson}) reduces to the standard Poisson equation of GR. This is partly due to the absence of a term like $Q(\dot\Phi+{\cal H}\Psi)$, such as occurred in the scalar-tensor and $f(R)$ cases. Terms like this originate from couplings between new degrees of freedom and the curvature scalar $R$ in the action. A direct coupling of this type isn't present in the Einstein-Aether case -- instead, the coupling between the aether and the metric is enforced through a Lagrange multiplier. However, it is not true that Einstein-Aether theory reduces to GR in the super-horizon regime, because the slip relation still has a non-GR form (eqn.(\ref{EA_USigma})).

\subsection{DGP gravity}
\label{subsection:DGP}

Dvali-Gabadadze-Porrati (DGP) gravity \cite{DGP2000} has received much attention over the past decade. The model considers our (3+1)-dimensional spacetime to be a hypersurface (brane) embedded in a five-dimensional bulk, with matter fields confined to the brane but gravity free to propagate into the extra dimension. There are two branches to the theory, arising from a choice of sign accompanying a square root in the Friedmann equation \cite{MaartensKoyama}:
\begin{equation}
\label{DGP_Friedmann}
{H}^2-\frac{\epsilon H}{r_c}=\frac{8\pi G}{3}\sum_i\rho_i 
\end{equation}
where $\epsilon=\pm 1$. Note that we have returned to using physical time rather than conformal time here in order to make the late-time behaviour more explicit. The parameter $r_c$ in this expression defines a crossover scale below which gravity is effectively four-dimensional, but above which five-dimensional effects become important. It is determined by the ratio of the four- and five-dimensional gravitational coupling constants:
\begin{equation}
\label{crossover_scale}
r_c=\frac{\kappa_5^2}{2\kappa_4^2}
\end{equation}
 In a CDM-dominated universe we have $H\rightarrow \epsilon/r_c$ at asymptotically late times. The choice $\epsilon = 1$ corresponds to a universe that accelerates without the need for a cosmological constant, although the crossover scale must still be fine-tuned to fit current supernovae data, with $r_c \approx H_0^{-1}$ \cite{MaartensKoyama}. Alas, the self-accelerating branch of DGP has been effectively ruled out as a viable theory due to a ghost-like instability \cite{DGP_ghost1, DGP_ghost2, DGP_ghost3}, and its failure to fit multiple observational data sets simultaneously -- a feat achieved by $\Lambda$CDM \cite{Fang_Wang_DGP}. 
Nevertheless, we propose to describe the relations between DGP gravity and the PPF formalism as a typical model for cosmological perturbations in higher-dimensional theories. 

We will adopt the Gaussian Normal longitudinal (GNL) gauge, in which the brane remains unperturbed at the hypersurface $y=0$ and we recover the familiar 4D conformal Newtonian gauge \textit{on the brane only}. The action for a simple DGP model is:
\begin{eqnarray}
\label{DGP_action} 
S &=& \frac{1}{2\kappa_5^2}\int d^5X \sqrt{-^{(5)}g}\, \left[^{(5)}R-2\Lambda_5\right]\nonumber\\
&&+\,\int d^4x\, \sqrt{-^{(4)}g}\left[\frac{1}{2\kappa_4^2}{^{(4)}R}+{\cal L}_B\right]
\end{eqnarray}
where $^{(5)}g$ and $^{(5)}R$ are the five-dimensional metric and curvature scalar of the bulk; $X^A$ and $x^{\mu}$ denote five-dimensional and brane coordinates respectively.
${\cal L}_B$ is the Lagrangian of the (brane-confined) matter fields and possible brane \makebox{tension $\lambda$}, which is related to the 5D and induced 4D cosmological constants \cite{Bridgman}.

It will be useful to write the modifications to the Friedmann and Raychaudhuri equations in the form of an `XY' background, as specified in eqns.(\ref{E_F}) and (\ref{E_R}) \cite{MG_report}:
\begin{eqnarray}
X&=&\frac{3\epsilon}{r_c}H\nonumber\\
Y&=&-\epsilon\frac{\left(\frac{dH}{dt}+3H^2\right)}{r_c H}
\end{eqnarray}
where t denotes physical time. Effective 4D field equations are obtained by projecting onto the brane \cite{Shiromizu}:
\begin{equation}
 \label{4D_DGP_einstein}
^{(4)}G_{\mu\nu}+\Lambda_4 g_{\mu\nu}=\kappa_5^4\Pi_{\mu\nu}-E_{\mu\nu}
\end{equation}
Here $g_{\mu\nu}$ is the induced metric on the brane. The tensor $\Pi_{\mu\nu}$ (not to be confused with the scalar pressure perturbation $\rho\Pi$) is given by:
\begin{equation}\label{DGP_Pi}
\Pi_{\mu\nu}=-\frac{1}{4}\tilde{T}_{\mu\alpha}\tilde{T}^{\alpha}_{\nu}+\frac{1}{12}\tilde{T}\tilde{T}_{\mu\nu}+\frac{1}{8}g_{\mu\nu}\tilde{T}_{\alpha\beta}\tilde{T}^{\alpha\beta}-\frac{1}{24}g_{\mu\nu}\tilde{T}^2
\end{equation} 
where
\begin{equation}
\label{Ttilde}
\tilde{T}_{\mu\nu}=T_{\mu\nu}-\frac{1}{\kappa_4^2}G_{\mu\nu}
\end{equation}
$E_{\mu\nu}$ is the projection of the 5D Weyl tensor onto the brane:
\begin{equation}
 E_{\mu\nu}=^{(5)}C^A_{BCD}n_An^Cg_{\mu}^B g_{\nu}^D     
 \end{equation}
By supplementing the Bianchi identities with the Codazzi equation and Israel junction conditions \cite{Israel_1966} one finds that the matter energy-momentum tensor is separately conserved from $\Pi_{\mu\nu}$ and $E_{\mu\nu}$ \,\cite{Shiromizu}: 
\begin{eqnarray}
\label{DGP_consv_1}
\nabla^{\mu}T_{\mu\nu}&=&0\\
\nabla^{\mu}E_{\mu\nu}&=&\kappa^4_5 \nabla^{\mu}\Pi_{\mu\nu}\label{DGP_consv_2}
\end{eqnarray}
where $\nabla^{\mu}$ is the covariant derivative associated with the induced 4D metric on the brane. 

Comparison of eqns.(\ref{einstein}) and (\ref{4D_DGP_einstein}) enables us to straightforwardly write down the tensor $U_{\mu\nu}$ for DGP gravity:
\begin{equation}
\label{DGP_U_background}
U_{\mu\nu}=-\Lambda_4 g_{\mu\nu}+\kappa_5^4\Pi_{\mu\nu}-E_{\mu\nu}-\kappa_4^2 T_{\mu\nu}
\end{equation}

Now the key question is: are we able to perturb this expression for $U_{\mu\nu}$ to obtain $U_{\Delta}, U_{\Theta}, U_P$ and $U_{\Sigma}$ in the same way that we have done for scalar-tensor gravity, $f(R)$ theories and Einstein-Aether theory (\textsection\ref{subsection:scalar_tensor}-\textsection\ref{subsection:EA})? The answer is a qualified yes and no. We \textit{are} able to write down expressions for these quantities and determine the coefficients laid out in \textsection\ref{section:extra_fields}. However, the perturbed components of $U_{\mu\nu}$ contain four new d.o.f. arising from the perturbations to the projected Weyl tensor. Hence, unlike the previous three examples presented in this section, the system is not closed using just the Bianchi identities.
We cannot obtain the closure relations needed to solve for these new d.o.f. using the purely 4D formalism developed in this paper.

Let us elaborate. We will label quantities related to the Weyl tensor by the letter E, as we shall see it plays the role of the effective fluid discussed in \textsection\ref{section:extra_fields}.
Since the Weyl tensor is traceless by construction (and we are assuming isotropy amongst the three spatial dimensions of the brane) we deduce that the Weyl `fluid' must have a radiation-like equation of state, $w_E=1/3$. It is common practice to neglect the small contribution of the Weyl fluid to the cosmological background, and define its components at the perturbed level:
\begin{eqnarray}
-E^0_0&=&\mu_E\nonumber\\
-E^0_i&=&\vec{\nabla}_i\Theta_E\nonumber\\
E^i_j&=&\frac{1}{3}\mu_E \delta^i_j+D^i_j \sigma_E
\end{eqnarray}
The perturbations of eqn.(\ref{DGP_U_background}) are then:
\begin{eqnarray}
\label{DGP_Us}
U_{\Delta}&=&\frac{1}{6}\kappa_5^4 a^2\tilde{\rho}^2\tilde{\delta}-a^2\mu_E-\kappa_4^2a^2\rho\delta \\
U_{\Theta}&=&\frac{1}{6}\kappa_5^4a^2 \tilde{\rho}\,(\tilde{\rho}+\tilde{P})\tilde{\theta}-a^2\Theta  -\kappa_4^2 a^2(\rho+P)\theta\nonumber\\
U_P&=&\frac{1}{2}\kappa_5^4 a^2[(\tilde{\rho}+\tilde{P})\tilde{\rho}\tilde{\delta}+\tilde{\rho}^2\,\tilde{\Pi}]-a^2\mu_E-3\kappa_4^2 a^2\rho\Pi\nonumber\\
U_{\Sigma}&=&-\frac{1}{12}\kappa_5^4 a^2(\tilde{\rho}+3\tilde{P})(\tilde{\rho}+\tilde{P})\tilde{\Sigma}-a^2\sigma_E\nonumber\\
&&\hspace{4.3cm}-\kappa_4^2 a^2(\rho+P)\Sigma\nonumber
\end{eqnarray}
The quantities marked by tildes are components of $\tilde{T}_{\mu\nu}$ defined in eqn.(\ref{Ttilde}), and the unsubscripted quantities refer to ordinary matter components, as in previous sections. By using eqns.(\ref{einstein}) and (\ref{Ttilde}) we can rewrite the above expressions as:
\begin{eqnarray}
\label{DGP_Us2}
U_{\Delta}&=&\frac{3}{2r_c^2 X}\left(E_{\Delta}+a^2\mu_E\right)\nonumber\\
U_{\Theta}&=&\frac{3}{2r_c^2 X}\left(E_{\Theta}+a^2\Theta_E\right)\nonumber\\
U_P&=&\frac{3}{2r_c^2 X}\Bigg(E_P-\frac{1}{2}a^2\mu_E\nonumber\\
&&-\frac{X r_c^2}{2Xr_c^2-3}(1+w_E)\left(E_{\Delta}-U_{\Delta}+a^2\mu_E\right)\Bigg)\nonumber\\
U_{\Sigma}&=&-\frac{3}{r_c^2 (X+3Y)}\left(E_{\Sigma}+a^2\sigma_E\right)
\end{eqnarray}
Well below the crossover scale we recover GR, as $U_i\rightarrow 0$ if $r_c\rightarrow\infty$. Using the definitions for the $E_i$, we can then put the above expressions into the form of eqns.(\ref{rescaled_general_Udelta2})-(\ref{extrafields_Sigma}). For the $U_{\Delta}$ component the non-zero coefficients are:
\begin{eqnarray}
A_0&=&-\frac{3}{r_c^2 X} \qquad A_1=-\frac{9{\cal H}_k}{r_c^2 X}\nonumber\\
F_0&=&-\frac{9{\cal H}^2_k}{r_c^2 X} \qquad \alpha_0^{(1)}=\frac{3a^2}{2 r_c^2 X}
\end{eqnarray}
where we have dropped the factor of $k^2$ in the first term of eqn.(\ref{rescaled_general_Udelta2}) in order to keep $\alpha_0^{(1)}$ dimensionless.
Unlike the previous examples, we now have a vector of three extra d.o.f.: $\vec{\chi}=\{\mu_E, \Theta_E, \sigma_E\}$.

The two perturbed components of eqn.(\ref{DGP_consv_2}) are \cite{MG_report}:
\begin{eqnarray}
\label{DGP_pert_consv1}
&&\dot{\mu}_E+4{\cal H}\mu_E-\nabla^2\Theta_E=0\\  
&&\dot{\Theta}_E+4{\cal H}\Theta_E-\frac{1}{3}\mu_E+(1+w_E)\left(\mu_E+3\frac{\cal H}{a}\Theta_E\right)\nonumber+\\
&&\frac{\nabla^2}{1+3w_E}\left(\frac{4}{3}\sigma_E+\frac{2(1+w_E)}{a^2}\left[(2+3 w_E)\Phi-\Psi)\right]\right)=0 \nonumber\label{DGP_pert_consv2}\\
\end{eqnarray}
Now the difficulty is apparent -- we have two equations for the three variables $\mu_E, \Theta_E$ and $\sigma_E$.
We could eliminate $\mu_E$ or $\Theta_E$ from the above equations, but the anisotropic stress of the Weyl fluid remains a free function. If we stay entirely within the bounds of a 4D formalism it must be treated as an additional source in the 4D effective Einstein equations. Some authors have obtained a closed system on the brane by setting $\sigma_E$ to zero, e.g. \cite{SawickiCarroll}, but this will not be the case is general.

It is not surprising that $\sigma_E$ appears as a free function in the perturbed effective Einstein equations. Gravitational waves propagating in the bulk spacetime contribute to $\sigma_E$ when they impinge upon the brane, and these cannot be fully described by brane-bound perturbations. However, by tackling the full system of perturbations in the bulk it is possible to express the impact of the gravitational waves on the brane in terms of other brane-bound quantities. Mukohyama \cite{Mukohyama} has shown that all 5D metric, matter and Weyl-fluid perturbations can be related to a master variable $\Omega$ which obeys a partial differential equation in the bulk. This is the master equation \cite{Deffayet2002}:
\begin{equation}
\label{master_equation}
\frac{\partial}{\partial t}\left(\frac{1}{na^3} \frac{\partial\Omega}{\partial t}\right)-\frac{\partial^i\partial_i}{a^2}\left(\frac{n\Omega}{a^3}\right)-\left(\frac{n\Omega^{\prime}}{a^3}\right)^{\prime}=0 
\end{equation}
where primes denote derivatives with respect to the bulk co-ordinate $y$, and we have assumed a Minkowski bulk. The perturbations of the Weyl fluid are given in terms of $\Omega$ by \cite{Deffayet2002}:
\begin{eqnarray}
\mu_E&=&-\frac{k^4}{3 a^5}\Omega\Bigg{|}_b \label{Weyl_perts_Omega_delta}\\
\Theta_E&=&\frac{k^2}{3a^4}\left(\frac{\partial\Omega}{\partial t}-H\Omega\right)\Bigg{|}_b\label{Weyl_perts_Omega_theta}\\
\sigma_E&=&-\frac{1}{6a^3}\bigg[3 \frac{\partial^2\Omega}{\partial t^2}-3H\frac{\partial\Omega}{\partial t}+\frac{k^2}{a^2}\Omega-\frac{3}{H}\frac{dH}{dt}\Omega^{\prime}\bigg]\Bigg{|}_b\label{Weyl_perts_Omega_sigma}
\end{eqnarray}
where a subscript $b$ denotes evaluation on the brane and k is the 3-momentum on the homogeneous background. Eqns.(\ref{master_equation}) and (\ref{Weyl_perts_Omega_sigma}) are the expressions we need to close the system of perturbations. We have introduced an extra variable, $\Omega$, but this has been compensated for by the addition of \textit{two} new equations. A solution of the master equation is then sufficient to fully determine the system of perturbations. The fact that all perturbations can be related to a single scalar is a result of the high degree of symmetry present in this system (a maximally symmetric brane in a maximally symmetric bulk); $\Omega$ is not generally believed to have a physical interpretation.

But we are not quite out of the woods yet. Though the system of equations is now closed, it is \textit{not} closed on the brane alone. Eqns.(\ref{master_equation}) and (\ref{Weyl_perts_Omega_sigma}) depend on derivatives of the master variable normal to the brane, and hence in general knowledge of the bulk perturbations will be required to solve the system. 
Koyama \& Maartens \cite{KoyamaMaartens_structureDGP} neatly sidestepped this issue by considering perturbations in the small-scale, quasistatic regime of a Minkowski bulk. This enabled them to make the approximation $|H\Omega^{\prime}|\ll k^2\Omega/a^2$ so that the troublesome terms can be neglected, as well as time derivatives. Under these assumptions one finds that the energy density and anisotropic stress perturbations of the Weyl fluid are simply related by $\mu_E=2k^2\sigma_E$. 
In the quasistatic regime the master equation then has an analytic solution \cite{KoyamaMaartens_structureDGP}:
\begin{equation}
\label{DGP_quasistatic_soln}
\Omega=C (1+H y)^{-\frac{k}{aH}}
\end{equation}
where $C$ is a constant, and regularity of the bulk perturbations has been used to eliminate a second possible solution. Solving for the Weyl fluid and metric perturbations, the modified Poisson equation and slip relation are:
\begin{eqnarray}
-2k^2\Phi&=&8\pi G_0 a^2 \sum_i \rho_i\Delta_i \left(1-\frac{1}{3\beta}\right)\label{DGP_QS_Poisson}\\
k^2\left(\Phi-\Psi\right)&=&8\pi\frac{G_0}{3\beta}a^2\sum_i\rho_i\Delta_i \label{DGP_QS_slip}\\ 
&&\hspace{-1.5cm}\mathrm{where}\quad \beta=1-2 r_c H\left(1+\frac{1}{3H^2}\frac{dH}{dt}\right)\nonumber
\end{eqnarray}
It is easy to see that our PPF function $\tilde g$ should be identified with \mbox{$\left(1-3\beta\right)^{-1}$}. Combining eqns.(\ref{DGP_QS_Poisson}) and (\ref{DGP_QS_slip}) to eliminate matter terms, we obtain:
\begin{equation}
\label{DGP_QS_slip2}
\Phi-\Psi=\frac{2}{1-3\beta}\Phi
\end{equation}
From this we can deduce the PPF function \mbox{$\zeta=2/(1-3\beta)$}. Alternatively, in the $(Q,\eta_{\mathrm{slip}})$ format we obtain: \mbox{$Q=\left(1-\frac{1}{3\beta}\right), \eta_{\mathrm{slip}}=(3\beta-1)(3\beta+1)^{-1}$}.

For scales larger than the quasistatic regime the derivatives normal to the brane cannot be ignored. 
Sawicki, Song and Hu were able to evolve large-scale modes numerically by implementing a scaling ansatz \cite{SawickiSongHu_DGPhorizon}:
\begin{equation}
\label{scaling_ansatz}
\Omega=A(p)a^p G(x)
\end{equation}
where $x$ is the distance from the brane in units of the causal horizon and $p$ is a constant. The authors took a trial value for $p$ and solved the system of equations iteratively to obtain subsequent corrections.  
They recovered the quasistatic solution of Koyama and Maartens and were able to calculate the behaviour of horizon-scale modes. However, the labour involved in obtaining these solutions is decidedly non-trivial. If we wish to constrain general classes of gravitational theories simultaneously we need a method that does not require detailed numerical evolution for each theory individually. 
This is the motivation behind the GDM-based approach we put forward in \textsection\ref{section:extra_fields}. In the case of DGP gravity the new d.o.f. are already written in the form of fluid perturbations, and we suggest that other theories may be amenable to a similar treatment. 

Returning briefly to the quasistatic limit of eqns.(\ref{DGP_QS_Poisson}) and (\ref{DGP_QS_slip}), we note that the master-variable route to obtaining a modified Poisson equation and a slip relation was decidedly different to that taken in \textsection\ref{subsection:scalar_tensor}-\ref{subsection:EA}. So it is interesting to find that we have reached the same conclusion in all four cases: the usual two-function PPF parameterization works well in the quasistatic limit, but for scales larger than this it no longer captures the full behaviour of the theories studied here.

We should not be particularly surprised by this conclusion. Given the different physical mechanisms employed by the four theories studied here, it is quite a remarkable feat that they can be mapped onto a common framework at all, in any limit -- let alone a framework simple enough to express the departures from GR using only two functions. It is not unexpected to find that the correspondence between parameterization and theory does not hold perfectly for all scales.

\section{Discussion}
\label{section:discussion}
The wide array of modified gravity theories now present in the literature renders the individual testing of theories time-consuming and impractical. There is also the possibility that none of the theories currently under consideration are correct, or that General Relativity remains valid for all length scales and environments. The Parameterized Post-Friedmann formalism  provides a useful framework for constraining deviations from General Relativity without recourse to a specific underlying theory of modified gravity. 

In this paper we have shown how the constraints of energy conservation, gauge invariance and second-order field equations can be used to reduce the components of a purely metric theory to two free functions. This process has highlighted some consistency conditions that arise when adopting a parameterized form for modifications to GR. We have put forward a general structure for the perturbed Einstein equations that should be applicable to any single-metric theory for which the gravitational action is built from only curvature invariants.
However, when additional degrees of freedom are introduced into a gravitational theory (such as extra fields), our ability to make precise statements about the form of the modifications to the Einstein equations is reduced.

Though a rich zoo of underlying physical mechanisms has been put forward, ultimately we are concerned with their effects on observables. The observables themselves (galaxies, CMB photons) are predominantly controlled by the evolution of matter perturbations. There are only a limited number of ways that a theory of modified gravity can affect the matter perturbations: for example, by changing the strength of gravitational coupling or the relationship between the metric potentials $\hat\Phi$ and $\hat\Psi$. But these effects are, to a certain extent, degenerate with the presence of a second fluid \cite{Kunz_Sapone}. A reduced gravitational coupling could be mimicked by pressure support from another fluid. A non-zero gravitational slip could be introduced by a fluid with non-negligible viscosity. Hence we suggest that it may be possible to treat the extra d.o.f. of a theory by an \textit{effective} fluid; Generalized Dark Matter (GDM) provides one method of doing this in the case of spin-0 perturbations. An effective fluid-type approach of this kind is already present in DGP.

Such treatment is not strictly necessary when two or fewer new degrees of freedom are present. But if we wish to keep our parameterization as general as possible, then we need a method to deal with more than two degrees of freedom. The GDM approach provides closure relations for such cases. The GDM parameters $\{w_E, c^2_{\mathrm{eff}}, c_{\mathrm{vis}}\}$ cannot be assigned the direct interpretation they possess for real fluids. But if constraints on these parameters favour bizarre-seeming values that would be unphysical for a real fluid than this could indicate their origin to be modified gravitational laws and not dark energy.

To date, the mapping of theories onto the PPF framework has only been computed explicitly for a small number of cases. In this paper we have added Einstein-Aether to this collection. In the quasistatic limit the two-function parameterization has always been found applicable, but it is possible that there are some classes of theories which cannot be reduced to such a simple format. In addition, we have found that on horizon scales the parameterization of the modified Poisson and slip relations no longer matches onto the underlying expressions exactly. These large scales are important for accurate calculation of how the ISW effect and matter power spectrum are affected by modified gravity theories. So should we give up on the goal of a unified parameterization?

Not at all. Consider the analogous problem in dark energy; the equation of state is commonly reduced to just two numbers via the CPL parameterization \cite{Chevallier_Polarski, Linder_CPL}:
\begin{equation}
\label{CPL}
w(a)=w_0+w_a (1-a)
\end{equation}
Whilst it is unlikely that a physically-motivated dark energy model will map neatly onto this expression, eqn.(\ref{CPL}) provides a useful way to obtain constraints on the expected behaviour of dark energy. Of course the approach is not ideal -- if the equation of state were to behave in a radically different way to our expectations then it would not be adequately described by \{$w_0, w_a$\}. But the form of eqn.(\ref{CPL}) has physical motivation,  and gives us a way to tackle large classes of models without specialising to a particular theory.

The PPF formalism should be viewed in a similar way. Eqns.(\ref{Poisson}) and (\ref{2nd_order_slip}) may not match up exactly to all theories on all scales, but they do provide a phenomenological way to search for the approximate signatures we expect modified gravity to leave. Similarly, our proposal of mapping additional degrees of freedom onto a GDM framework may not hold exactly for all possible theories.
But whilst we have not yet reached an ideal parameterization of modified gravity, we believe that the approach outlined in this paper is likely to have a wider range of applicability than most of the forms currently in common use. If, when combined with the next generation of experiments, we fail to make progress on breaking the degeneracy between dark energy and modified gravity, then we will need to rethink our tools. For the present, use of the parameterizations described here is a justifiable simplification, provided that we are not lulled into thinking that their correspondence to underlying theories is exact in all situations.

If we are to use such a phenomenological methodology we need to choose a parameterization that closely models our expectations. Small differences in the structure of the modifications to Einstein's equations can lead to significant effects, independent of the ansatz being used for the PPF functions. We will demonstrate these differences and how they affect the constraints obtained in a future work \cite{Zuntz2011}.

\vspace{-10pt}
\section*{Acknowledgements}
\vspace{-10pt}
We are grateful for useful discussions with R. Bean, E. Bertschinger, L. Pogosian, A. Silvestri, D. Wands, C. Will and T. Zlosnik. TB is supported by the STFC.  PGF acknowledges support from the STFC, the Beecroft Institute for Particle Astrophysics and Cosmology, and the Oxford Martin School. CS is supported by a Royal Society University Research Fellowship. JZ is supported by the Dennis Sciama fellowship and a James Martin fellowship.

\appendix
\section{Gauge-Invariant Equations for Scalar-Tensor Theory }
\label{app:gauge_invariance}
In this Appendix we display explicitly the gauge-invariant form of the modifications to the perturbed Einstein equations in the case of scalar-tensor gravity. At the level of the homogeneous, background universe the modifications are contained in the diagonal components of a tensor $U_{\mu\nu}$, see eqn.(\ref{einstein}). The perturbed components of this tensor appearing in eqns.(\ref{einstein1})-(\ref{einstein4}) are denoted by $U_{\Delta}, U_{\Theta}, U_P$ and $U_{\Sigma}$ as given in eqns.(\ref{U_components_def}). In \textsection\ref{subsection:scalar_tensor} these components were derived in the conformal Newtonian gauge for simplicity. Using the notation introduced in \textsection\ref{section:formalism}, the expressions in a general gauge are:
\begin{widetext}
\begin{eqnarray}
U_{\Delta} &=& E_{\Delta} (1-\phi) +3\dot\phi\,\hat\Gamma+\left(\delta\phi-\dot\phi\frac{V}{2}\right) \left[\frac{1}{2}\frac{\mathrm{d}\omega
(\phi)}{\mathrm{d}\phi}\,\frac{\dot\phi^2}{\phi}-\frac{1}{2}\omega(\phi)\frac{\dot\phi^2}{\phi^2}-k^2-3 {\cal H}^2\right]\nonumber\\
&&+\left(\delta\dot\phi+\dot\phi\, \Xi+\frac{V}{2}({\cal H}\dot\phi-\ddot\phi)\right)\left[\omega(\phi)\frac{\dot\phi}{\phi}-3{\cal H}\right]
-\frac{3}{2}V {\cal H}\left[\omega(\phi)\frac{\dot\phi^2}{\phi}+\ddot\phi-2{\cal H}\dot\phi\right]\label{gi_st_1}\\
U_{\Theta}&=&E_{\Theta}(1-\phi)+\left(\delta\phi-\dot\phi\frac{V}{2}\right)\left[\omega(\phi)\frac{\dot\phi}{\phi}-{\cal H}\right]+\left(\dot{\delta\phi}+\dot\phi\,\Xi+\frac{V}{2}({\cal H}\dot\phi-\ddot\phi)\right)+\frac{V}{2}\left[\omega(\phi)\frac{\dot\phi^2}{\phi}+\ddot\phi-2{\cal H}\phi\right]\label{gi_st_2}\\
U_P &=& E_P (1-\phi)-6\dot\phi\,\hat\Gamma
+\,3\left(\delta\phi-\dot\phi\frac{V}{2}\right)\left[-\frac{1}{2}\omega(\phi)\frac{\dot\phi^3}{\phi^2}+\frac{1}{2} \frac{\mathrm{d}\omega(\phi)}{\mathrm{d}\phi}\,\frac{\dot\phi^2}{\phi}+{\cal H}^2+2\dot{\cal H}+\frac{2}{3}k^2\right]\nonumber\\
&&+3\left(\dot{\delta\phi}+\dot\phi\,\Xi+\frac{V}{2}({\cal H}\dot\phi-\ddot\phi)\right)\left[{\cal H}+\omega(\phi)\frac{\dot\phi}{\phi}\right]
+\,3\left(\delta\ddot\phi +2\Xi\,(\ddot\phi-{\cal H}\dot\phi)+\dot\phi(\dot\Xi+{\cal H}\Xi)+\frac{V}{2}\left(2{\cal H}\ddot\phi-E\frac{\dot\phi}{2}-\phi^{(3)}\right)\right)\nonumber\\
&&+\,\frac{3}{2}V\phi\left[\frac{\partial}{\partial\tau}\left(\omega(\phi)\frac{\dot\phi^2}{\phi^2}+\frac{\ddot\phi}{\phi}+{\cal H}\frac{\dot\phi}{\phi}\right)-2{\cal H}\left(\omega(\phi)\frac{\dot\phi^2}{\phi^2}+\frac{\ddot\phi}{\phi}+\,{\cal H}\frac{\dot\phi}{\phi}\right)\right]\label{gi_st_3}\\
U_\Sigma &=&E_{\Sigma}(1-\phi)+\left(\delta\phi-\dot\phi\frac{V}{2}\right)\label{gi_st_4}
\end{eqnarray}
\end{widetext}
In these expressions the scalar field perturbations $\delta\phi, \delta\dot\phi$ and $\delta\ddot\phi$ appear only in gauge-invariant combinations with the variables $V$ and $\Xi$. $\hat\Gamma$ is also gauge-invariant. The last terms in  $U_{\Delta}, U_{\Theta}$ and $U_P$ are not gauge-invariant, and neither are  $E_{\Delta}, E_{\Theta}$ and $E_P$. But we find that the additional terms produced by these parts under a gauge transformation cancel by virtue of the background equations (\ref{st_background_1}) and (\ref{st_background_2}) - see Table \ref{tab:gauge_transf} for the transformation properties. Note that all terms in the spatial, traceless Einstein equation are individually gauge-invariant.

So the perturbed Einstein equations with the additions above remain fully gauge-invariant, as they must do. However, on first inspection it looks like we have violated the constraint of having second-order field equations. Recall that this constraint restricts $U_{\Delta}$ and $U_{\Theta}$ to contain at first-order time derivatives at most, due to eqns.(\ref{Bianchi1}) and (\ref{Bianchi2}). It is easy to see that the $\ddot\phi$ terms present in $U_{\Theta}$ cancel, but in $U_{\Delta}$ this is not explicitly obvious. In addition, $U_{\Delta}$ appears to contain a second-order time derivative of the scale factor coming from the $\dot{\hat\Phi}$ term (see eqn.(\ref{phi_def})). To show this is not a problem we will need to use the zero-order Bianchi identity, eqn.(\ref{E_Bianchi}). In the case of scalar-tensor theory this becomes:
\begin{equation}
\omega(\phi)\left(\frac{\ddot\phi}{\phi}-\frac{1}{2}\frac{\dot\phi^2}{\phi^2}+2{\cal H}\frac{\dot\phi}{\phi}	\right	)+\frac{1}{2}\frac{\mathrm{d}\omega}{\mathrm{d}\phi}\frac{\dot\phi^2}{\phi}-3\left(\dot{\cal H}+{\cal H}^2\right)=0 \label{st_background_Bianchi}
\end{equation}
where we have made use of eqn.(\ref{st_background_1}). Using this to substitute for the $\omega\ddot\phi$ term in eqn.(\ref{gi_st_1}) and simplifying we obtain the alternative form:
\begin{eqnarray}
\label{st_udelta_alternative}
U_{\Delta}=&&E_{\Delta}(1-\phi)-\frac{1}{2}\dot\phi\dot J-3{\cal H}\dot\phi\Xi-\delta\phi\left(k^2+3{\cal H}^2\right)\nonumber\\
&&+\frac{1}{2}k^2\dot\phi\,V+\delta\phi\left[\frac{1}{2}\frac{\mathrm{d}\omega}{\mathrm{d}\phi} \frac{\dot\phi^2}{\phi}-\frac{1}{2}\omega(\phi)\frac{\dot\phi^2}{\phi}\right]\nonumber\\
&&+\left(\delta\dot\phi+\dot\phi\Xi\right)\left[\omega(\phi)\frac{\dot\phi}{\phi}-3{\cal H}\right]
\end{eqnarray}
We see that the terms containing second derivatives of $\phi$ and the scale factor have both cancelled. The above expression is now demonstrably first-order in time derivatives, but its ability to yield gauge-invariant Einstein equations is no longer obvious.

\section{Constraint Equations}
\label{app:constraints}
In the case of unmodified background equations, constraints are obtained by applying the Bianchi identity to a $U$-tensor of the general format shown in eqns.(\ref{Udelta})-(\ref{Usigma}). It is possible to deduce the general structure of these constraint equations and use this as a shortcut. The constraint equations arising from the coefficient of $\Phi^{(n)}$ can be generated from the following formulae:
\begin{eqnarray}
&[B1]& \; \dot A_n+k A_{n-1}+{\cal H}A_n+k B_n+{\cal H} C_n = 0 \label{B1_Phi_constraint} \\
&[B2]& \; \dot B_n+k B_{n-1}+2 {\cal H} B_n-\frac{1}{3} k C_n+\frac{2}{3}k D_n =0 \label{B2_Phi_constraint}
\end{eqnarray}
where $B1$ indicates eqn.(\ref{Bianchi1}), $B2$ indicates eqn.(\ref{Bianchi2}). 
The constraints arising from the term $\hat\Gamma^{(n)}$ are analogous:
\begin{eqnarray}
&[B1]& \; \dot F_n+k F_{n-1}+{\cal H}F_n+k I_n+{\cal H} J_n = 0 \label{B1_Psi_constraint} \\
&[B2]& \; \dot I_n+k I_{n-1}+2 {\cal H} I_n-\frac{1}{3} k J_n+\frac{2}{3}k K_n =0 \label{B2_Psi_constraint}
\end{eqnarray}
Note that for an Nth-order theory, $F_{N-2}$ and $I_{N-2}$ must be set to zero to avoid picking up (N+1)-th derivatives of the scale factor.


\bibliographystyle{apsrev}

\end{document}